\documentclass[10pt,twocolumn,letterpaper]{article}

\usepackage{iccv}
\usepackage{times}
\usepackage{epsfig}
\usepackage{graphicx}
\usepackage{amsmath}
\usepackage{amssymb}

\usepackage{eso-pic}
\usepackage{multirow}
\usepackage{setspace}
\usepackage{comment}
\usepackage{siunitx}
\usepackage{adjustbox}
\usepackage{booktabs}
\usepackage{caption} 
\usepackage[dvipsnames]{xcolor}
\usepackage{arydshln}
\usepackage{indentfirst}
\usepackage{xcolor}
\usepackage[T1]{fontenc}
\usepackage[symbol]{footmisc}
\usepackage{wrapfig,lipsum,booktabs}
\usepackage{caption}

\usepackage{ifthen}
\newboolean{sepline} 
\setboolean{sepline}{true} 

\usepackage[breaklinks=true,bookmarks=false]{hyperref}

\iccvfinalcopy 

\def\iccvPaperID{6633} 

\ificcvfinal\fi
 
\begin{document}

\title{Efficient Unified Demosaicing for Bayer and Non-Bayer Patterned Image Sensors}

\author{Haechang Lee$^{1,4,*}$, \ \ Dongwon Park$^{2,*}$, \ \ Wongi Jeong$^{1,*}$, \\ 
Kijeong Kim$^{4}$, \ \ Hyunwoo Je$^{4}$, \ \ Dongil Ryu$^{4}$ \ and \ Se Young Chun$^{1,2,3,\dagger}$\\
$^1$Dept. of ECE, \ $^2$INMC, \ $^3$IPAI, \ Seoul National University, \ Republic of Korea,\\
$^4$SK hynix, \ Republic of Korea\\
{\tt\small \{harrylee,dong1park,wg7139,sychun\}@snu.ac.kr}
}

\maketitle

\let\thefootnote\relax\footnotetext{$*$  Equal contribution, $\dagger$ Corresponding author.}

\begin{abstract}
As the physical size of recent CMOS image sensors (CIS) gets smaller, the latest mobile cameras are adopting unique non-Bayer color filter array (CFA) patterns (e.g., Quad, Nona, Q$\times$Q), which consist of homogeneous color units with adjacent pixels. These non-Bayer sensors are superior to conventional Bayer CFA thanks to their changeable pixel-bin sizes for different light conditions, but may introduce visual artifacts during demosaicing due to their inherent pixel pattern structures and sensor hardware characteristics. Previous demosaicing methods have primarily focused on Bayer CFA, necessitating distinct reconstruction methods for non-Bayer patterned CIS with various CFA modes under different lighting conditions. In this work, we propose an efficient unified demosaicing method that can be applied to both conventional Bayer RAW and various non-Bayer CFAs' RAW data in different operation modes. Our Knowledge Learning-based demosaicing model for Adaptive Patterns, namely KLAP, utilizes CFA-adaptive filters for only 1\% key filters in the network for each CFA, but still manages to effectively demosaic all the CFAs, yielding comparable performance to the large-scale models. Furthermore, by employing meta-learning during inference (KLAP-M), our model is able to eliminate unknown sensor-generic artifacts in real RAW data, effectively bridging the gap between synthetic images and real sensor RAW. Our KLAP and KLAP-M methods achieved state-of-the-art demosaicing performance in both synthetic and real RAW data of Bayer and non-Bayer CFAs.
\end{abstract}

\begin{figure*}
    \centering
    \includegraphics[width=0.95\textwidth]{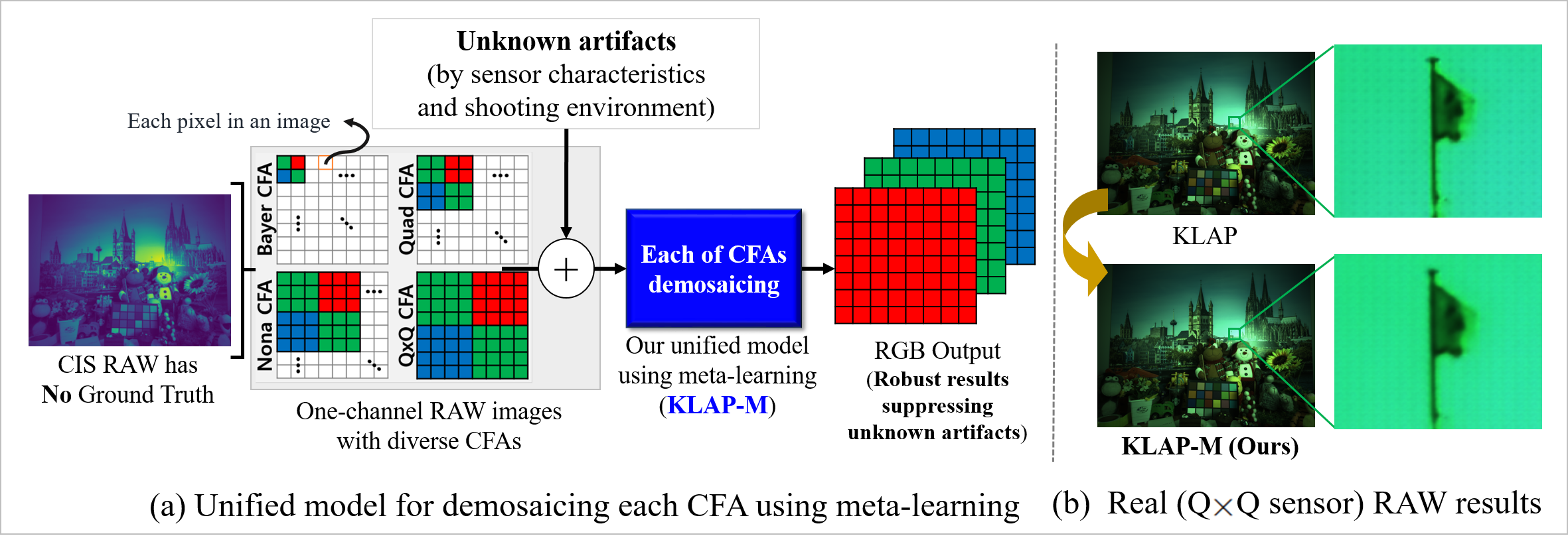}
    \vspace{-1em}
    \caption{(a) Overview of our unified model (UM) for demosaicing all the Bayer and non-Bayer CFAs, called the \textbf{K}nowledge \textbf{L}earning-based demosaicing model for \textbf{A}daptive \textbf{P}atterns using \textbf{M}eta-test learning (KLAP-M) with Bayer or non-Bayer patterns, even when ground truth is unavailable and unknown artifacts are present. (b) Comparing CIS RAW demosaicing results of KLAP (KLAP-M without meta-test learning) and KLAP-M (KLAP with meta-test learning).}
    \label{Fig1_overview}
        \vspace{-1em}
\end{figure*}

\vspace{-1em}

\section{Introduction}
\label{sec:intro}
Demosaicing (DM) is the process of interpolating single-channel input images into RGB output images within an embedded Image Signal Processor (ISP). With the growing demand for high-quality mobile camera images, CMOS image sensor (CIS) resolution has increased dramatically, even reaching 200 million pixels in the latest smartphones. However, as image sensors cannot infinitely increase in size, pixel size has been reduced to enhance resolution.
Smaller CISs are more vulnerable to noise and degradation in image restoration capabilities because they are more sensitive to variations in light reception, especially in low-light condition~\cite{diamond2021dirty,jung20221, oh20200, okawa20191}.
As a result, modern high-end smartphones have started using image sensors that group adjacent homogeneous pixels, resulting in non-Bayer Quad, Nona, and Quad-by-Quad (Q$\times$Q) sensors~\cite{jang20190, oh20200, park2022advanced}, while still retaining some of the properties of the standard Bayer CFA~\cite{bayer1976color} pattern.
Quad, Nona, and Q$\times$Q sensors combine the same color pixel arrays of 2$\times$2, 3$\times$3, and 4$\times$4 respectively, resulting in homogeneous pixel units (\textit{i.e.}, Gr, R, B, and Gb) for each sensor, as shown in Fig.~\ref{Fig1_overview}(a).

Demosaicing for modern non-Bayer CFAs is more complex and computationally demanding than for standard Bayer CFAs. This is because as the number of pixel arrays within each unit increases, the distance between the units becomes greater, requiring interpolation with inaccurate pixel values from distant locations.
Therefore, there is growing interest in using deep learning for demosaicing methods, leading to active research on both Bayer pattern demosaicing~\cite{zhang2022deep,pistellato2022deep,dewil2023video,chen2021joint,xing2021end,a2021beyond,liu2020joint,zamir2020cycleisp,schwartz2018deepisp,ignatov2020replacing,kim2020pynet,ignatov2022pynet} and non-Bayer pattern demosaicing~\cite{kim2019deep, kim2021recent, jeong2022color, arad2022ntire, sharif2021sagan, cho2022pynet}.

However, the aforementioned methods focus on a single CFA pattern task and do not cover demosaicing tasks for other CFA patterns. Modern mobile phones with non-Bayer patterned CIS adapt their CFA modes dynamically based on lighting conditions, controlled by the CIS's ISP.
Using independent models (IMs) for each pattern, tailored to different CFA modes, would demand loading and operating multiple models within the limited circuit space of the CIS. This would result in excessive memory and power consumption if the models were kept standby on the mobile application processor (AP) and switched accordingly. Moreover, the task of tuning models for each CFA would be laborious.

Currently, no existing method can handle dynamically changing CFA modes in a non-Bayer patterned CIS as a unified model (UM).
Inspired by recent works for all-in-one image restoration affected by multiple types of unknown degradation~\cite{li2020all,chen2022learning,li2022all}, we propose a unified demosaicing method for all Bayer and non-Bayer CFA patterns.
However, these all-in-one image restoration methods do not consider real unknown artifacts, so we will further investigate them to address the scenario of real CIS RAW with `unknown' artifacts, missing or mostly lacking ground truth (GT). Since such unknown artifacts may fail to yield high-quality phone camera photos, we are motivated to propose a UM with robust meta-learning-based DM methods that can handle these obstacles.

In this work, we propose efficient unified DM methods that are capable of handling various non-Bayer patterned CISs with a new pipeline to bridge the gap between synthetic and real CIS RAW images.
Our proposed \textbf{K}nowledge \textbf{L}earning-based demosaicing model for \textbf{A}daptive \textbf{P}atterns (KLAP) is capable of simultaneously handling multiple CFAs' demosaicing, which consists of two following steps.
Firstly, we train a baseline UM using the two-stage knowledge learning (TKL)~\cite{chen2022learning}, making it more efficient to find Adaptive Discriminative filters for each specific CFA Pattern (ADP).
Secondly, we fine-tune the UM model trained in the first stage using ADP. ADP is a metric that applies FAIG~\cite{xie2021finding} to the update logic of our neural network, allowing us to find a small set of discriminative filters that can be used as independent parameters for specific CFA DM tasks.
Lastly, we propose KLAP-M, KLAP (TKL+ADP) with Meta-test learning. KLAP-M integrates self-supervised learning into KLAP to address domain gaps between synthetic RAW and real CIS RAW images caused by unknown artifacts in real-life scenarios.
Our proposed meta-test learning for demosaicing consists of pixel binning loss based on CIS domain knowledge and self-supervised denoising techniques.
Fig.~\ref{Fig1_overview}(a) provides an overview of our KLAP-M approach, which handles both Bayer and Non-Bayer patterns. Additionally, Fig.~\ref{Fig1_overview}(b) shows the results of our meta-test learning technique, addressing the domain gap in real RAW images.

Our contributions are summarized as follows: (1) Our efficient unified network, KLAP, effectively performs demosaicing for multiple CFAs, (2) KLAP-M, a version of KLAP that incorporates a meta-learning approach, effectively reduces unknown visual artifacts in genuine CIS RAW images that are caused by diverse sensor characteristics and shooting environments, (3) KLAP and KLAP-M achieve state-of-the-art performance on the synthetic benchmark dataset and real CIS RAW samples captured by CIS chips.

\section{Related Works}
\subsection{Deep Learning-based Demosaicing}
\label{sec:DL_DM}

\textbf{IMs for DM only.} Traditional demosaicing without applying deep learning techniques either apply a fixed DM filter to each pixel without considering other parameters as features or utilize spectral and spatial features available in neighboring pixels to interpolate the unknown pixel as closely as possible to the original~\cite{mahajan2015adaptive,heide2014flexisp}.
Due to the complexity of various CIS CFAs, traditional methods are cumbersome, leading to an increasing interest in deep learning-based demosaicing models.
Stojkovic \textit{et al.}~\cite{stojkovic2019effect} suggested IMs of each Bayer and Quad demosaicing based on CDMNet~\cite{cui2018color}.
Kim \textit{et al.}~\cite{kim2019deep, kim2021recent} applied the duplex pyramid network structure to Quad CFA and Nona CFA, respectively.
Sharif \textit{et al.}~\cite{sharif2021sagan} proposed a GAN-based spatial-asymmetric attention for Nona CFA reconstruction.
For Q$\times$Q CFA, Cho \textit{et al.}~\cite{cho2022pynet} proposed an efficient pyramidal network using progressive distillation based on PyNet~\cite{ignatov2020replacing}.

\textbf{Multi-tasks joint with DM.} There have been proposals to combine DM methods with other closely related ISP tasks, such as denoising (DN) and super-resolution (SR).
Some~\cite{dewil2023video, wu2023learning, chen2021joint, liu2020joint, jin2020review, kokkinos2018deep} proposed convolutional neural networks approach for joint DM and DN to improve the quality of the restored image. Ma \textit{et al.}~\cite{ma2022searching} and Xu \textit{et al.}~\cite{xu2020joint} proposed models for simultaneous DM and SR. Xing \textit{et al.}~\cite{xing2021end} introduced a multi-task learning approach to jointly address three tasks: DM, DN, and SR.
Previous studies mainly concentrate on multi-task approaches for single CFA demosaicing and known noise sources. In contrast, our proposed method introduces a unified model that handles both Bayer and non-Bayer CFAs, incorporating meta-learning to ensure robust performance even in the presence of unknown noise.

\subsection{Image Restoration for Multi-tasks}
\label{sec: IR for Multi-tasks}
\textbf{IMs for multi-tasks.} Beyond DM tasks, recent papers~\cite{zamir2021multi, mou2022deep, zamir2022restormer, wang2022uformer, tu2022maxim, purohit2021spatially, chen2022simple} have introduced various approaches that share a common framework capable of addressing multiple image restoration tasks, including denoising, deblurring, and deraining.
While the mentioned IM excels in individual tasks, it necessitates multiple network parameters as multiple networks are needed to handle all the required tasks.

\textbf{Unified model (UM) for multi-tasks.} To overcome the drawbacks of IMs, Chen et al~\cite{chen2022learning} proposed a single UM for two-stage knowledge learning mechanism based on multi-teacher and single student approach for multiple degradations on images that contains rain, haze, and snow.
Li \textit{et al.}~\cite{li2022all} proposed a single UM using a contrastive-based degraded encoder, called the degradation-guided restoration network (DGRN), which adaptively works with three degradations: rain, noise, and blur.
Park \textit{et al.}~\cite{park2023all} introduced a single UM equipped with dedicated filters for degradation, achieving remarkable results in rain-noise-blur and rain-snow-haze tasks.
To the best of our knowledge, there is currently no reported method that can handle all Bayer and non-Bayer demosaicing tasks using a single unified model.

\subsection{Meta-learning-based Image Restoration}
For image reconstruction, a large number of samples are usually necessary, but it may not be feasible in many real-world situations. Meta-learning, also known as learn-to-learn, provides a promising solution to the problem of adapting models quickly to new data. This learning method empowers models to achieve efficient task performance even with limited additional incoming data. Finn \textit{et al.}~\cite{finn2017model} proposed an algorithm for model-agonistic meta-learning that achieved state-of-the-art performance in few-shot learning tasks. Meta-SR~\cite{hu2019meta} enables super-resolution for arbitrary scale factors by applying the Meta-Upscale Module.
We propose the use of meta-learning to achieve robust results, even in the presence of unknown artifacts in CIS RAW images.

\section{Deep Demosaicing for Each Non-Bayer CFA}
\subsection{Operating Principles of Non-Bayer Sensors} 
\label{sec:OperatingNonBayer}

\begin{figure}[!b]
    \vspace{-1.5em}
    \centering
    \includegraphics[width=0.995\linewidth]{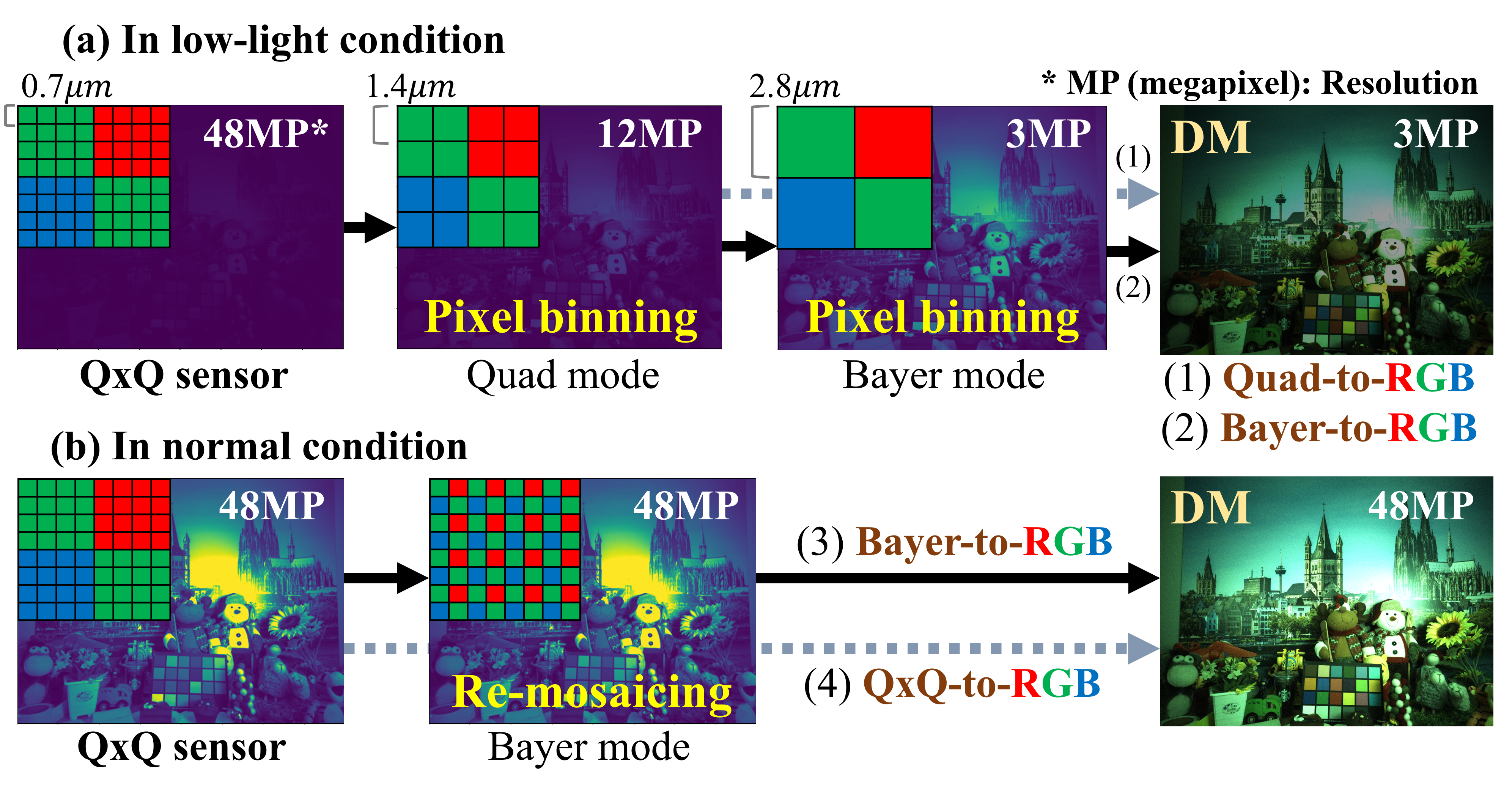}
    \vspace{-1.5em}
    \caption{DM scenario in real CIS. For example, in the case of Q$\times$Q CIS:
    (a) In low-light conditions, the Q$\times$Q sensor converts its pattern to either the (1) Quad or (2) Bayer mode (pixel-binning), sacrificing resolution, and then performs DM.
    (b) In normal conditions, the Q$\times$Q sensor can either re-mosaic the pattern to the Bayer mode and then perform DM or directly perform Q$\times$Q  DM, with full resolution.}
    \label{Fig2_DMscenario}
\end{figure}

With the decreasing size of camera sensors, the physical area of light captured by a pixel has been reduced. Consequently, the introduction of non-Bayer sensors allows for capturing more light. In case of Q$\times$Q as an example, when there is sufficient light, as scenario (3) and (4) in Fig.~\ref{Fig2_DMscenario}, Q$\times$Q sensors can handle the entire resolution with Bayer DM (after `re-mosaicing') and direct Q$\times$Q DM.
On the other hand, especially in low-light conditions, Q$\times$Q CIS pixels have the advantage of using `pixel binning' to enhance their light sensitivity and reduce the noise~\cite{zhou1997frame, yoo2015low}, sacrificing their resolution (but still acceptable), resulting in clear image quality with reduced noise (shown as scenario (1) and (2) in Fig.~\ref{Fig2_DMscenario}). Pixel binning is the merging of neighboring pixels in an image through summation or averaging in ISP, typically done by the ISP after pixel-readout. Quad DM or Bayer DM methods are specifically required in such cases.
Supporting a diverse range of CFA pattern modes remains crucial in non-Bayer patterns. However, employing separate DM networks for each pattern increases network parameters, leading to larger CIS chip area. Multiple DM models necessitate frequent model switching, consuming more memory and power in mobile environment.
Our proposed unified DM model handles all non-Bayer sensor patterns, including standard Bayer sensors, providing effective solutions for this issue. It offers flexibility for different CIS product lines and CFA pattern modes, reducing product development time with minimal fine-tuning required for specific product characteristics.

\subsection{Data Synthesis for Demosaicing All CFAs} 
\label{sec:Synthesis}

\label{sec:synthesis_pipeline}
\begin{figure}[!ht]
    \centering
    \includegraphics[width=0.98\linewidth]{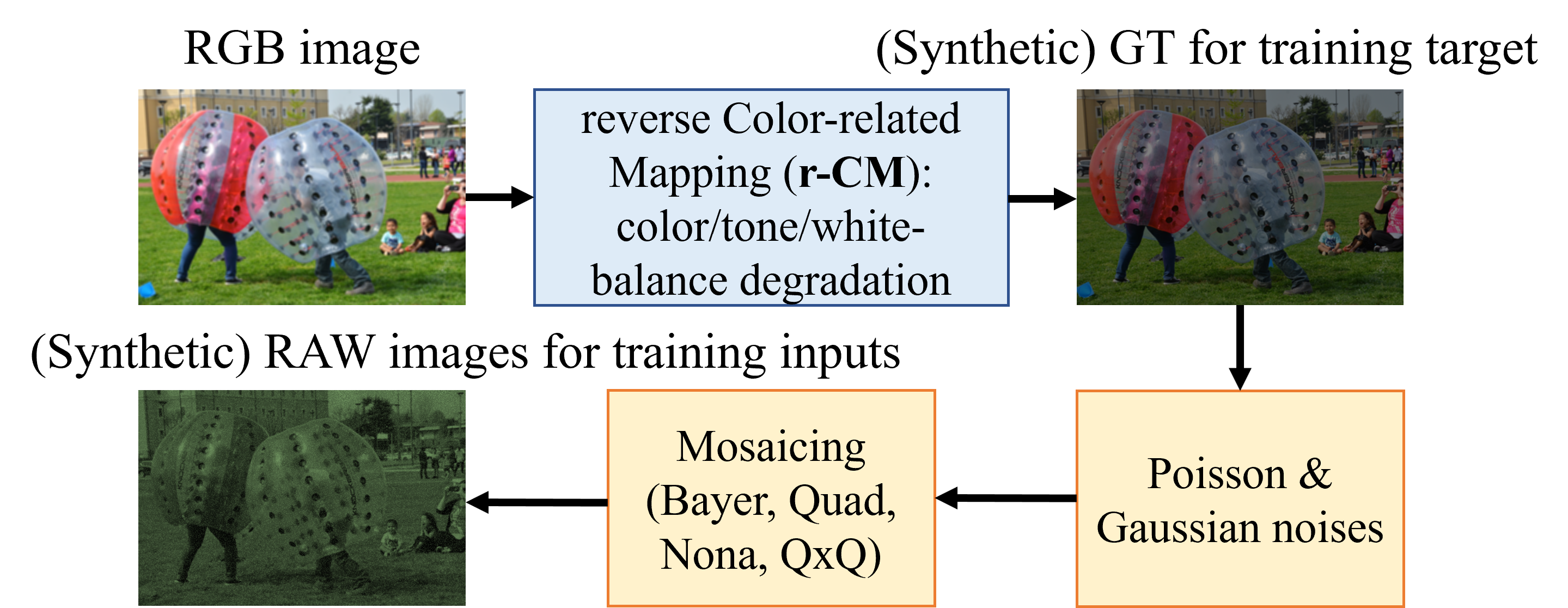}
    \caption{Overview of our realistic RAW image synthesis pipeline for Bayer and Non-Bayer demosaicing.  
    The r-CM (reverse Color-related Mapping functions) towards RAW-like synthesis consists of invertible linear operations that relate RGB color spaces.
    }
    \label{Fig3_data_synthesis}
\vspace{-1.5em}
\end{figure}

To train input images resembling real CIS RAW, we propose a data synthesis pipeline that generates realistic RAW-like images. Using a high-quality sRGB dataset, we follow the front-end of Fig.~\ref{Fig3_data_synthesis} to generate synthetic RAW-like images. This involves applying four reverse color-related mapping functions (r-CM) from the ISP chain, including color tone degradation, inverse gamma correction, inverse color correction, and inverse auto white balance correction functions.
We analyzed and adjusted the previous ISP chains, resulting in a pipeline structure similar to previous methods.~\cite{tseng2022neural,tseng2021differentiable,rim2022realistic,brooks2019unprocessing,xu2020noisy}.
Using this method, we generate RGB synthetic GT labels for demosaicing training.
Furthermore, we add Gaussian and Poisson noise to simulate various types of real noise~\cite{rim2022realistic, diamond2021dirty, brooks2019unprocessing, xu2020noisy}. Each image is then converted into a mosaic pattern for Bayer, Quad, Nona, and Q$\times$Q CFA, as depicted in the bottom row of Fig.~\ref{Fig3_data_synthesis}. This process generates the training inputs.
The reverse color mapping (r-CM) consists of linear operations and can be easily "re-reversed" to obtain the original color mapping (CM). CM makes final output images only after DM that closely resemble human-viewed realistic images.
Our proposed synthetic dataset generation pipeline considers demosaicing for both Bayer and Non-Bayer patterns and incorporates a realistic noise model that combines Gaussian and Poisson noise.
More detailed information is Sec. S.1 in the supplementary material.

\subsection{Domain Gap in Synthetic and Real CIS RAW} 
\label{sec:domainGap}

Synthetic data-trained models often struggle with real data due to the domain gap issue, a persistent problem in image restoration tasks~\cite{rim2022realistic, brooks2019unprocessing, ji2020real}.
The domain gap arises from variations in sensor hardware characteristics due to differences in circuit structure, manufacturing processes, and component variations across CIS brands and product lines. The upper image in Fig.~\ref{Fig1_overview}(b) shows visual artifacts in real CIS RAW, mainly caused by crosstalk effects~\cite{khabir2021electrical, khabir2022characterization, li2022crosstalk} between inner and outer pixels (details in Sec. S.2 in the supplementary). Moreover, unknown artifacts can emerge in different shooting environments and vary across CIS types. To address this, we propose a meta-learning method to minimize the domain gap, enabling the effective handling of unexpected unknown artifacts.

\begin{figure*}
    \centering
    \includegraphics[width=0.96\textwidth]{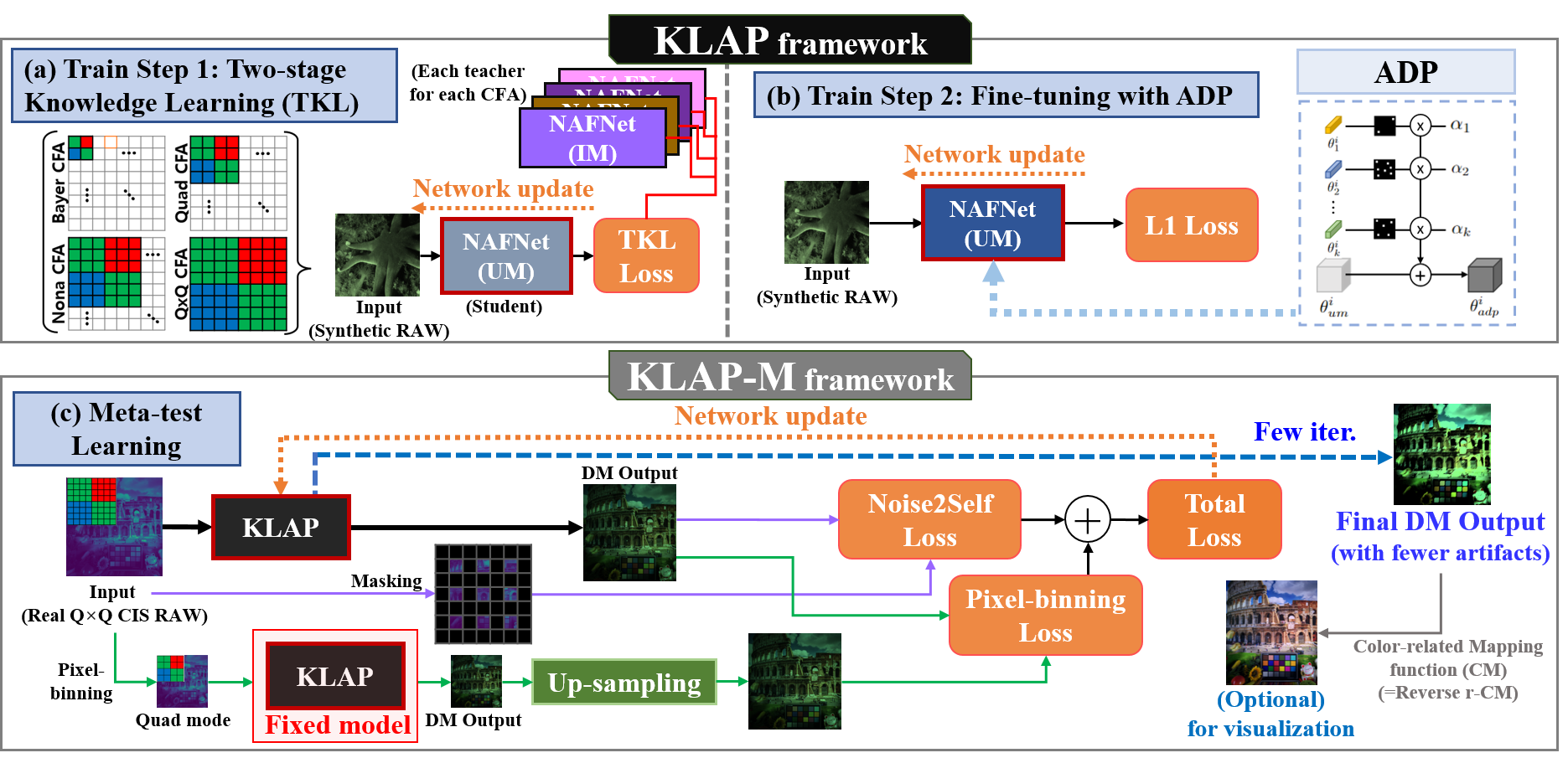}
    \caption{The overview of our proposed unified DM model, Knowledge Learning-based demosaicing model for Adaptive Pattern (KLAP) and KLAP with Meta-test learning (KLAP-M). KLAP consists of 2 steps: (a) two-stage knowledge learning (TKL) for training baselines, (b) fine-tuning using Adaptive Discriminant filters for each specific CFA Pattern (ADP). (c) KLAP-M employs meta-learning to reduce unknown artifacts in real RAW images during inference.
    }
    \label{Fig4_KLAP-M_arch}
\end{figure*}

\section{Unified Deep Demosaicing for Multiple Bayer and Non-Bayer CFAs}
\label{methods}
Fig.~\ref{Fig4_KLAP-M_arch} displays the proposed single unified DM method for all Bayer and non-Bayer sensor patterns (KLAP) and its additional meta-learning during inference framework for robustness (KLAP-M). In Step 1 as Fig.~\ref{Fig4_KLAP-M_arch}(a), our approach augments the network capacity of the integrated model using the Two-stage Knowledge Learning~\cite{chen2022learning} (TKL). This maximize the effectiveness of the Adaptive Discriminative filters for each specific CFA Pattern (ADP) discovered in the subsequent step. In Step 2 as Fig.~\ref{Fig4_KLAP-M_arch}(b), we further enhance the UM using a small number of specialized network kernels for each DM task. Lastly, as Fig.~\ref{Fig4_KLAP-M_arch}(c), we introduce a meta-test learning framework that ensures robust DM output in the presence of unknown artifacts.
 
\subsection{Step 1: Two-stage Knowledge Learning}
\label{TKL}
This step aims to train the baseline of unified DM model (baseline UM) for all CFAs using the two-stage knowledge learning~\cite{chen2022learning} (TKL), with independent DM models for each CFA (IMs). The IMs, with the same network architecture, have independent network parameters ($\{\theta_i\}_{i=1}^k$) dependent on each CFA-specific DM task ($k$). Note that the IM achieves high performance as a specialized model for each task, but requires a model k times larger than UM ($\theta_{um}$).

First, we pre-train each individual IM based on NAFNet~\cite{chen2022simple}, renowned for its high performance despite having few network parameters. Then, in the knowledge collection (KC) stage, set the IMs specialized for each CFA DM task as the teacher network and UM as the student network to learn and collect knowledge from the teacher. In the knowledge examination (KE) stage after KC, train only using the student network and GT labels without guidance from the teacher network.
We applied TKL method to increase the model's capacity after feature-level guidance for each CFA pattern, in order to maximize the effect of top filter detection in FAIG (Filter Attribution method based on Integral Gradient)~\cite{xie2021finding} (see actual results in Tab.~\ref{tab:AblationStudy}).

\subsection{Step 2: Adaptive Discriminative Filters for a specific CFA Pattern}
\label{sec:ADP}
Xie \textit{et al.}~\cite{xie2021finding} proposed FAIG, which can detect discriminative filters of specific degradation. FAIG measures integrated gradient (IG)~\cite{sundararajan2016gradients, sundararajan2017axiomatic} between baseline and target models. Inspired by FAIG and its application in another domain~\cite{park2023all}, we applied CNN for Adaptive Discriminant filters for a specific CFA Pattern (ADP) using the leveraged FAIG method. FAIG score is as follows : $FS_j = FAIG_j(\theta_{um}, \theta_i, x_i),$ for multiple CFA filters $i=1,\ldots,k$ and all kernels $j$. Once the FAIG scores are computed, they are then ranked in descending order. The top $q$\% of kernels are selected for each demosaicing process, with $q$ representing a fixed value between 0.5 and 5.

We propose ADP, implemented by the masks $M_{c}$ that are selected kernels using FAIG as illustrated in Fig.~\ref{Fig4_KLAP-M_arch} and defined as follows:
\begin{equation}
\theta^i_{adp} = \theta^i_{um} + \sum_{c=1}^{k}{~ \alpha_c \theta^{i}_{c} * M^{i}_{c} }
\end{equation}
where $i$ is kernel index, $*$ is point-wise multiplication, $\theta^i_{um}$ refers to the pre-trained integrated model in Step 1, and $\alpha_{c}$ is a coefficient for a specific CFA pattern and is set up either as 1 or 0. Note that in a real non-Bayer CIS on a mobile device, the pattern mode $i$ is determined by the mobile AP after detecting the lighting conditions. Also $\theta^{i}_{c}$ is an additional kernel for specific CFA pattern. The ratio $q$ in the mask is determined empirically to be $1\%$.
For example, the ratio of $1\%$ in the mask is $1\%$ for 4 demosaicing types, our proposed method uses an additional $4\%$ of the entire network parameters as compared to the baseline UM. More detailed information is in the supplementary Sec. S.3. Our proposed KLAP, a combination of TKL and ADP, achieves state-of-the-art performance in various CFA DM tasks by replacing only relevant CNN kernels in UM from TKL.

\subsection{Meta-learning during Inference}
\label{sec:Meta-learning}
As shown in Fig.~\ref{Fig4_KLAP-M_arch}, we propose meta-learning during inference (meta-test learning) to mitigate unknown artifacts caused by sensor characteristics or shooting environments. By performing a few network updates during inference, this approach produces robust results.
Our proposed meta-learning during inference consists of pixel binning loss and Noise2Self (N2S) loss, one of the self-supervised denoising techniques.
As mentioned in Sec.~\ref{sec:OperatingNonBayer}, pixel binning compensates for resolution loss by increasing the light sensitivity, thus reducing noise. Based on CIS domain knowledge, we propose a self-supervised denoising method using a pixel binning loss to remove unknown artifacts.
\begin{equation}
\mathcal{L}_\text{pix} = \left\vert G(x_{J^c},\theta_{adp}) - U(G(m(x_{J^c}),\theta_{adp'})) \right\vert,
\end{equation}
where $x$ and $J^c$ denote the CIS RAW data and mask used by N2S, $m$ and $U$ represent average-based pixel-binning operation and up-sampling operation, respectively. $G$ is a unified network structure and $\theta_{adp}$ is network parameters of ADP. $\theta_{adp'}$ is the initial network parameters that are not updated.  

Additionally, we apply modified N2S loss to maintain robustness against noise (Poisson and Gaussian noise) that may occur depending on the shooting environment and to prevent blur caused by pixel binning loss:
\begin{equation}
\mathcal{L}_\text{N2S} = \left\vert G(x_{J^c},\theta_{adp})_J - x_J \right\vert
\end{equation}
where $x_J$ and $x_J^c$ are represent independent images using the mask scheme. Additional information about the pixel binning loss and N2S loss can be found in the supplementary material (See Sec. S.4.2.)

The total loss for meta-learning during inference is as follows:
\begin{equation}
\mathcal{L}_\text{total} = \lambda_\mathrm{pix} \mathcal{L}_\mathrm{pix} + \lambda_\mathrm{N2S} \mathcal{L}_\mathrm{N2S}
\end{equation}
where $\lambda_\mathrm{pix}$ and $\lambda_\mathrm{N2S}$ are used to balance different loss conditions and is experimentally found through visualization.

\begin{figure*}[!ht]
\centering
\includegraphics[width=0.80\textwidth]{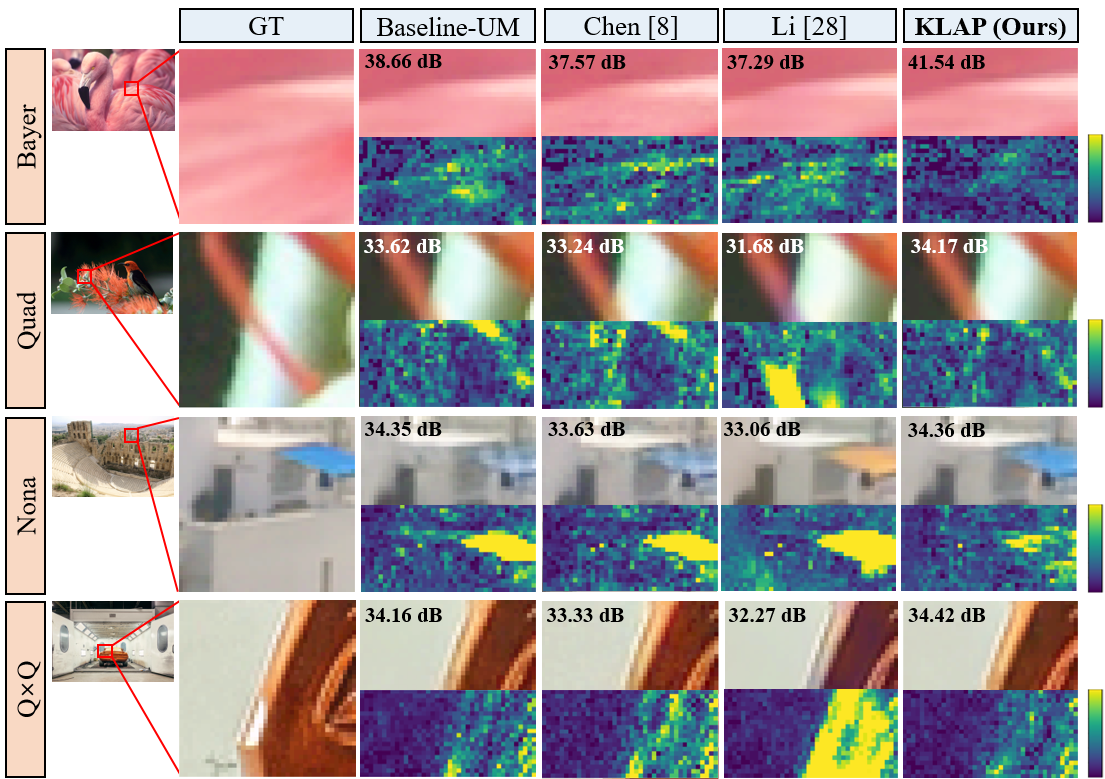}
    \caption{
    Comparisons of demosaiced images {\bfseries(top)} from different methods and their difference maps {\bfseries(bottom)} on the synthetic RAW (DF2K-CIS) test dataset. The PSNR (dB) value displayed in the top-left corner is for the entire image. 
    As shown in the figure above, our proposed KLAP achieves the best performance in synthetic RAW test dataset.
    }
    \label{Fig6_DF2K-CIS}
\end{figure*}

\section{Experimental Results} 
\label{sec:Datasets}
As stated in Sec.~\ref{sec:Synthesis}, we generate synthetic DF2K Bayer and Non-Bayer CIS (DF2K-CIS) dataset utilizing DF2K, a combination of two open source datasets, DIV2K~\cite{agustsson2017ntire} and Flickr2K~\cite{Lim_2017_CVPR_Workshops}.
The training set comprises 2,500 images, with a validation set of 50 images and a test set of 1000 images. Furthermore, we propose to use the DF2K-CIS test dataset with strong noise to evaluate the effectiveness of our proposed meta-test learning in generating robust results.
The DF2K-CIS strong noise test dataset comprises 200 images with noise parameters four times larger than those used in training. Then, we evaluate our proposed meta-learning method, KLAP-M, using 7 Q$\times$Q CIS RAW images (48MP) with a resolution of $8000\times 6000$, 
1 Quad CIS RAW image, and 3 Bayer CIS RAW images (50MP) with a resolution of $8192\times 6144$, all of which are 10-bit images captured directly by each type of CIS chip. In the meta-test learning, KLAP-M is trained using the loss function in Eq.(4) with $\lambda_{pix} = 1$ and $\lambda_{N2S} = 0.02$. Note that Meta-test (KLAP-M) does not utilize IMs but instead employs a unified model, and we conducted KLAP-M evaluations on each new full image for each sensor type.
More implementation details and demosaicing RAW results can be found in Sec. S.5, S.7 and S.8. of the supplementary materials

\begin{table}[!t]
\caption{Ablation study for our proposed KLAP and Quantitative performance comparison (Chen~\cite{chen2022learning} and Li~\cite{li2022all}) on DF2K-CIS test dataset in terms of PSNR (dB) and the number of parameters (Million). 
 Baseline-UM is a simple unified model. TKL is Baseline UM applying TKL, and ADP is Baseline UM applying ADP independently. TKL-to-IM involves fine-tuning IM after applying TKL. Chen~\cite{chen2022learning} and Li~\cite{li2022all} are based on MSBDN~\cite{dong2020multi} and AirNet, respectively, while other experiments are based on NAFNet~\cite{chen2022simple}.
Note that Avg. denotes mean of all CFA's PSNR, and Par. denotes the number of parameters.}
\label{tab:AblationStudy}
\centering
\begin{adjustbox}{width=0.98\linewidth}
\begin{tabular}{c|cccc|c|c}
\hline
Method  & Ba. & Qu. & No. & QxQ & Avg. & Par. \\ \hline\hline
IM & 42.18 & 41.80 & 41.14 & 41.42 & 41.64 & \textcolor{red}{\textbf{68.4}}          \\
TKL-to-IM & \textbf{42.36} & \textbf{41.89} & \textbf{41.58} & \textbf{41.60} & \textbf{41.86} & \textcolor{red}{\textbf{68.4}} \\\hline
Baseline UM & 41.90   &  41.40 & 41.03   & 41.09  &  41.35 & 17.1  \\
Baseline UM-L & 41.95 &  41.44 & 41.08  &  41.13 &  41.40  &  19.4 \\
Chen~\cite{chen2022learning} & 41.43 & 40.89 & 40.54 & 40.49 & 40.84 & 28.7          \\
Li \cite{li2022all} & 38.28 & 38.08 & 38.23 & 36.94 & 37.88   & \textbf{7.6}           \\ \hline
TKL & 41.89  & 41.44  & 41.11  & 41.15  & 41.40  &  17.1 \\
ADP & 42.06  & 41.50  & 41.14  & 41.16  & 41.46  & 17.8  \\ \hline
KLAP (Ours) & \textbf{42.25}  & \textbf{41.75} & \textbf{41.42} & \textbf{41.41}  & \textbf{41.71} & \textcolor{blue}{\textbf{17.8}} \\ \hline
\end{tabular}
\end{adjustbox}
    \label{tab:ablation}
\vspace{-1em}
\end{table}

\begin{figure}[!b]
    \vspace{-1em}
    \centering
    \includegraphics[width=0.47\textwidth]{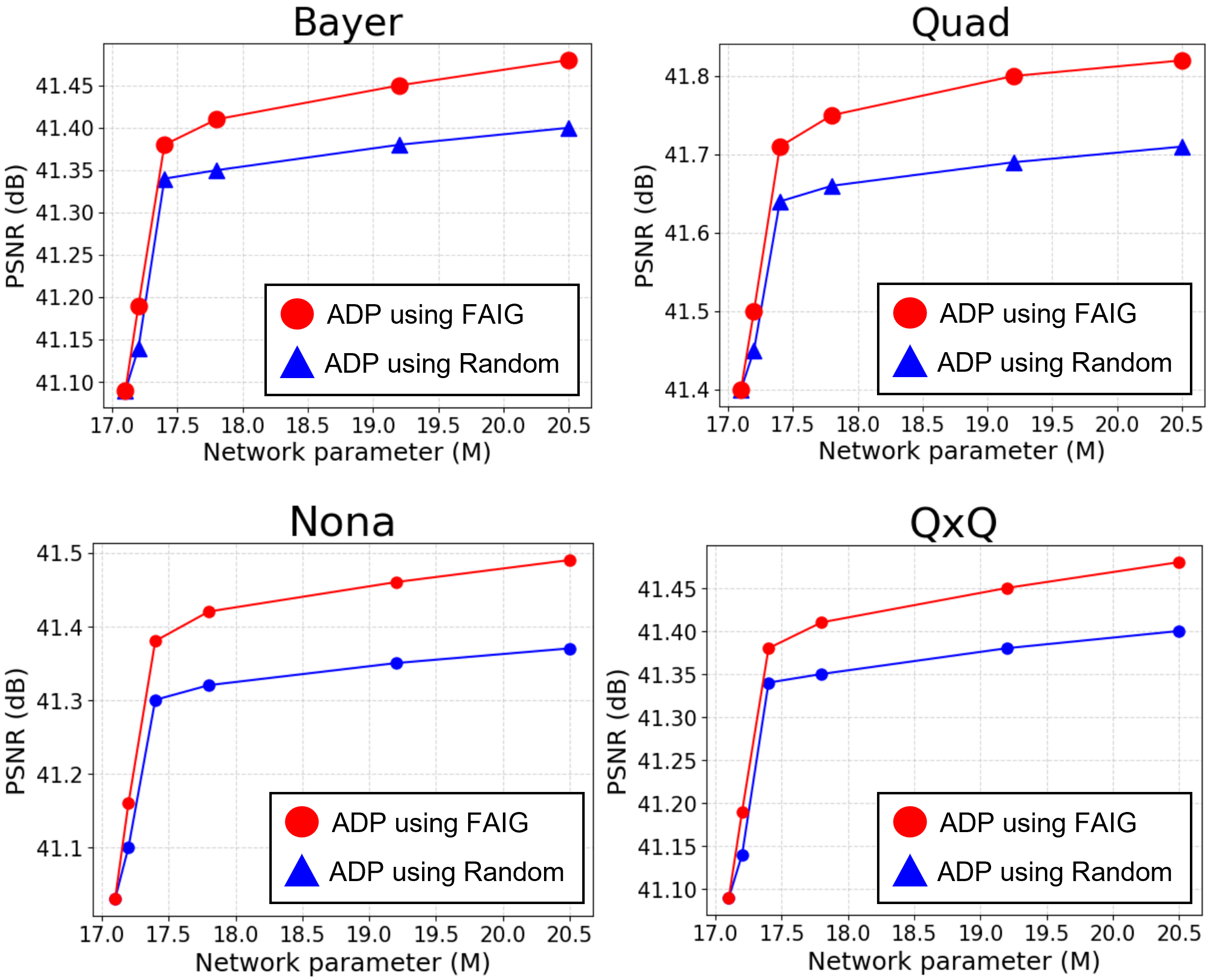}
    \caption{Performance comparisons among different filter location selections ($0\%$, $0.1\%$, $0.5\%$, $1\%$, $3\%$, and $5\%$, respectively) for UM with ADP: Random selection method and FAIG  adjusting ADP on DF2K CIS test dataset.}
\label{fig:selected_filter}
\vspace{-1em}
\end{figure}

\subsection{Results on Synthetic RAW Dataset}
\subsubsection{Comparison of Ablation Studies and KLAP with Other Methods}

\textbf{Ablation study for KLAP.}
We perform ablation studies on the proposed KLAP approach based on NAFNet~\cite{chen2022simple}, including TKL and ADP, as shown in Fig.~\ref{Fig4_KLAP-M_arch} (a) and (b), using the DF2K-CIS test dataset.
Tab.~\ref{tab:ablation} summarized the performance of PSNR (dB) and the number of parameters (Million). Baseline UM is a simple integrated model trained on all tasks, while total IMs require 4 times more network parameters than UM.
Baseline UM-Large (Baseline UM-L) refers to a modified version of NAFNet~\cite{chen2022simple} with increased network blocks. In TKL and ADP in the table, each step is independently applied to the baseline UM. TKL-to-IM refers to the re-trained IMs after applying TKL.

Using TKL and ADP independently leads to only a marginal improvement of 0.05 dB and 0.11 dB, respectively, compared to Baseline UM.
Our proposed KLAP (TKL+ADP) further improved performance by 0.4 dB with a slightly increased number of network parameters compared to Baseline UM.
Notably, Our KLAP achieved significantly higher performance than Baseline UM-L (41.71dB vs. 41.40dB) with fewer parameters (17.8M vs. 19.4M). In addition, fine-tuning each IM with pre-trained TKL resulted in a notable improvement compared to the original IMs, attributed to the inclusion of contrastive learning loss in TKL.
Our proposed KLAP method, which combines TKL and ADP, significantly improves demosaicing performance for all CFAs.

\textbf{Comparisons among other unifying methods.}
We evaluate the performance of our KLAP with NAFNet~\cite{chen2022simple} on a DF2K-CIS test dataset and summarize the results in Tab.~\ref{tab:ablation} in terms of PSNR (dB) and the number of parameters.
We use the official codes provided by the authors of Airnet~\cite{li2022all} and Chen~\cite{chen2022learning}. The Chen~\cite{chen2022learning} method uses the MSBDN-based TKL method.
Despite a slight increase in network parameters by 0.7M (about 4\%) in NAFNet, our KLAP yields significantly improved performance by 0.4 dB compared to the IM method.
Notably, our KLAP yields the highest PSNR among all-in-one methods~\cite{chen2022learning,li2022all} while using smaller network parameters compared to existing methods applied to NAFNet networks. 
Fig.~\ref{Fig6_DF2K-CIS} shows DM results on synthetic datasets for visual comparisons. We adjust CM in Sec.~\ref{sec:synthesis_pipeline} for visualization. The images on the 1st to 4th rows are input synthetic RAW images and their DM outputs of UM, Chen~\cite{chen2022learning}, Li~\cite{li2022all}, and our KLAP are on the 2nd, 3rd, 4th,
5th column of Fig.~\ref{Fig6_DF2K-CIS}, respectively. This shows that our KLAP outperforms other state-of-the-art unifying methods on DF2K-CIS test datasets.

\begin{figure*}[!t]
\centering
\includegraphics[width=0.85\textwidth]{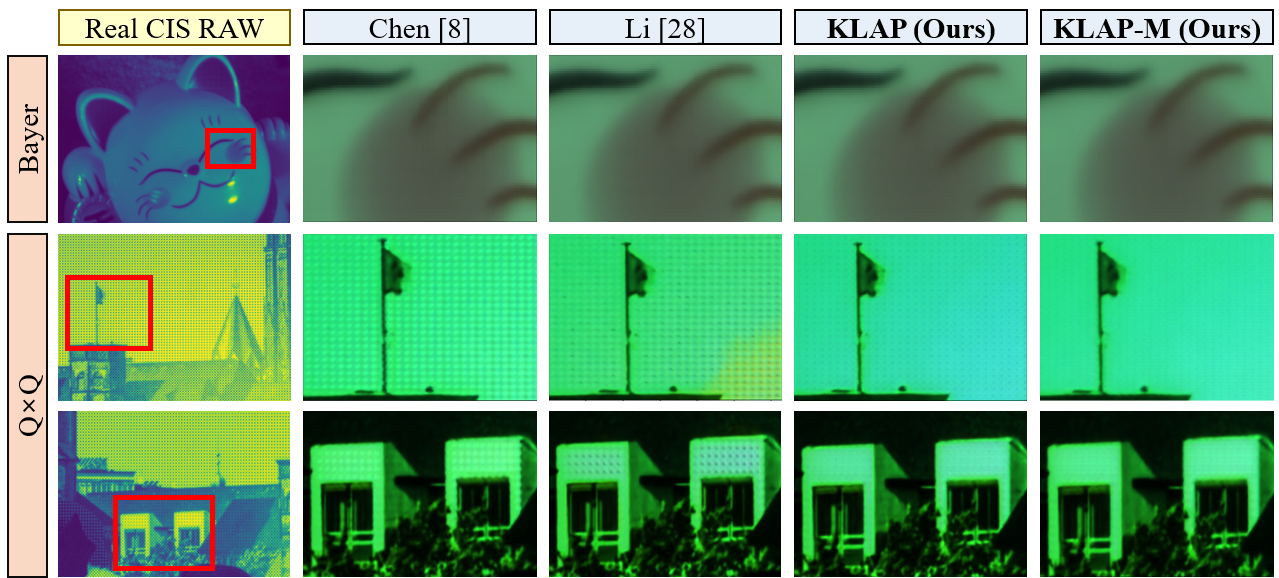}
    \caption{Qualitative DM results on the real CIS RAW. Note that KLAP with meta-test learning (KLAP-M) shows robust performance in real CIS RAW, despite of existence of sensor-generic unknown artifacts.
    }
    \label{Fig7_real_CIS}
\end{figure*}

\subsubsection{Performance and Selected Filter Locations}
To demonstrate the superiority of FAIG~\cite{xie2021finding} over random selection, we evaluate various mask selection strategies in our ADP method on synthetic datasets with both Bayer and non-Bayer patterns. The mask selection ratios are set to $0.1\%$, $0.5\%$, $1\%$, $3\%$, and $5\%$. We use a UM with TKL-based NAFNet~\cite{chen2022simple} and add adaptive network kernels in proportion to the $q$ ratio. Two mask selection methods are investigated: random selection and the FAIG method introduced in Sec.~\ref{sec:ADP}. 
Fig.~\ref{fig:selected_filter} summarizes our results, indicating that our ADP adopting FAIG outperforms random filter selection, underscoring the effectiveness of selecting discriminative filters for each CFA DM task. This implies that discriminative filters can be defined as task-specific (in our case, each CFA DM) filters, rather than randomly selected filters.

\subsubsection{Analysis of Robustness in Strong Noise}
To validate the robustness of Meta-test learning in KLAP-M, we evaluate KLAP-M on the DF2K-CIS with strong noise test dataset and summarize the results in a table. The DF2K-CIS with strong noise dataset has four times larger noise parameters compared to the DF2K-CIS training dataset. As shown in Tab.~\ref{tab:strong_noise}, KLAP shows slightly more robust results compared to existing methods. Furthermore, when KLAP-M is applied, it achieves an average improvement of 1.8 dB in PSNR with only 10 iterations.

\begin{table}[]
\caption{Performance comparisons among different methods of robustness with strong noise in terms of PSNR (dB) on DF2K-CIS test dataset with strong noise. The noise parameters used in the test are four times larger than the noise parameters used in the training, and the number of meta-learning iterations in KLAP-M is fixed to 10.
}
\centering
\begin{adjustbox}{width=0.90\linewidth}
\label{tab:strong_noise}
\begin{tabular}{c|ccc|cc}
\hline
CFA & Chen~\cite{chen2022learning} & Li~\cite{li2022all} & KLAP & KLAP-M  \\ \hline \hline
Bayer & 32.60 & 31.61 & 32.98 & \textbf{33.32} \\
Quad & 32.48 & 31.58 & 32.93 & \textbf{35.41} \\
Nona & 32.44 & 31.64 & 32.88 & \textbf{35.06} \\
Q$\times$ Q & 32.45 & 31.38 & 32.86 & \textbf{35.41} \\\hline
\end{tabular}
\end{adjustbox}
\vspace{-1em}
\end{table}

\begin{figure}
    \centering
    \includegraphics[width=0.47\textwidth]{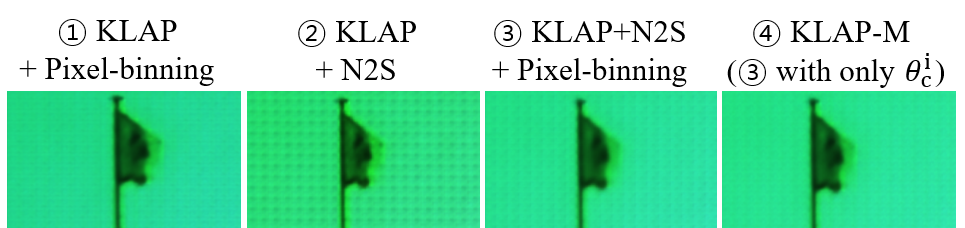}
    \caption{Ablation study of our proposed KLAP-M. The comparison shows the effect of each component of meta-learning in KLAP-M.}
\label{Fig8_real_CIS_ablation}
\vspace{-1.5em}
\end{figure}

\subsection{Results on Real CIS RAW}
We evaluate the performance of our KLAP with meta-learning on a real RAW dataset and present the results in Figure~\ref{Fig7_real_CIS}. The number of iterations for meta-learning is fixed at 45. In the Bayer case, our method, as well as Chen~\cite{chen2022learning} and Li~\cite{li2022all}'s methods, show robust results on real data.
However, In the case of demosaicing Q$\times$Q, Chen and Li's methods are unable to alleviate artifacts, while our method significantly mitigates resulting artifacts during inference by reducing domain gap through meta-learning. Figure~\ref{Fig8_real_CIS_ablation} shows the ablation study of KLAP-M and demonstrates superior performance compared to other method combinations. Note that the Bayer output is an image that has been squared by 0.7 from the original outputs for visual comparison purposes. We represent the two Q$\times$Q output images, with their pixel values (range of 0 to 1) cubed, to compare the artifact mitigation performance with other models.

\subsection{Limitations}
\label{sec:limitations}
To utilize deep learning-based DM models for CIS, the requirement of a specialized circuit with embedded AI accelerators can be a limiting factor.

\section{Conclusion}
\label{sec:conclusion}
Our proposed demosaicing method uses task-specific kernels to cover all CFAs and incorporates a meta-testing framework to produce efficient and robust results. This approach boasts low computational complexity, robustness to unknown artifacts, and high-quality demosaiced images.
\\

\noindent\textbf{Acknowledgments} This work was supported in part by the National Research Foundation of Korea(NRF) grants funded by the Korea government(MSIT) (NRF-2022R1A4A1030579), Basic Science Research Program through the NRF funded by the Ministry of Education(NRF-2017R1D1A1B05035810) and Creative-Pioneering Researchers Program through Seoul National University. The CIS RAW data and CIS domain knowledge were supported by CIS Development Representative at SK hynix.

{\small
\bibliographystyle{ieee_fullname}
\bibliography{egbib}
}


\clearpage
\appendix

\twocolumn[{%
\renewcommand\twocolumn[1][]{#1}%

\begin{center}
\bigskip 
\bigskip 
\textbf{\Large Efficient Unified Demosaicing for Bayer and Non-Bayer Patterned Image Sensors \\ (Supplementary Material) \\}
\bigskip 
\bigskip 
{\large Haechang Lee$^{1,4,*}$, \ \ Dongwon Park$^{2,*}$, \ \ Wongi Jeong$^{1,*}$, \\ 
Kijeong Kim$^{4}$, \ \ Hyunwoo Je$^{4}$, \ \ Dongil Ryu$^{4}$ \ and \ Se Young Chun$^{1,2,3,\dagger}$\\
$^1$Dept. of ECE, \ $^2$INMC, \ $^3$IPAI, \ Seoul National University, \ Republic of Korea,\\
$^4$SK hynix, \ Republic of Korea\\
{\tt\small \{harrylee,dong1park,wg7139,sychun\}@snu.ac.kr}
}
\bigskip 
\bigskip 

\maketitle

\setcounter{equation}{0} 
\setcounter{figure}{0} 
\setcounter{table}{0} 
\setcounter{page}{1} 
\makeatletter 

\renewcommand{\thefigure}{S.\arabic{figure}}
\renewcommand{\thetable}{S.\arabic{table}}
\renewcommand{\thesection}{S.\arabic{section}}

    \vspace{-1em}
    \centering
    \captionsetup{type=figure}
    \includegraphics[width=0.8\textwidth]{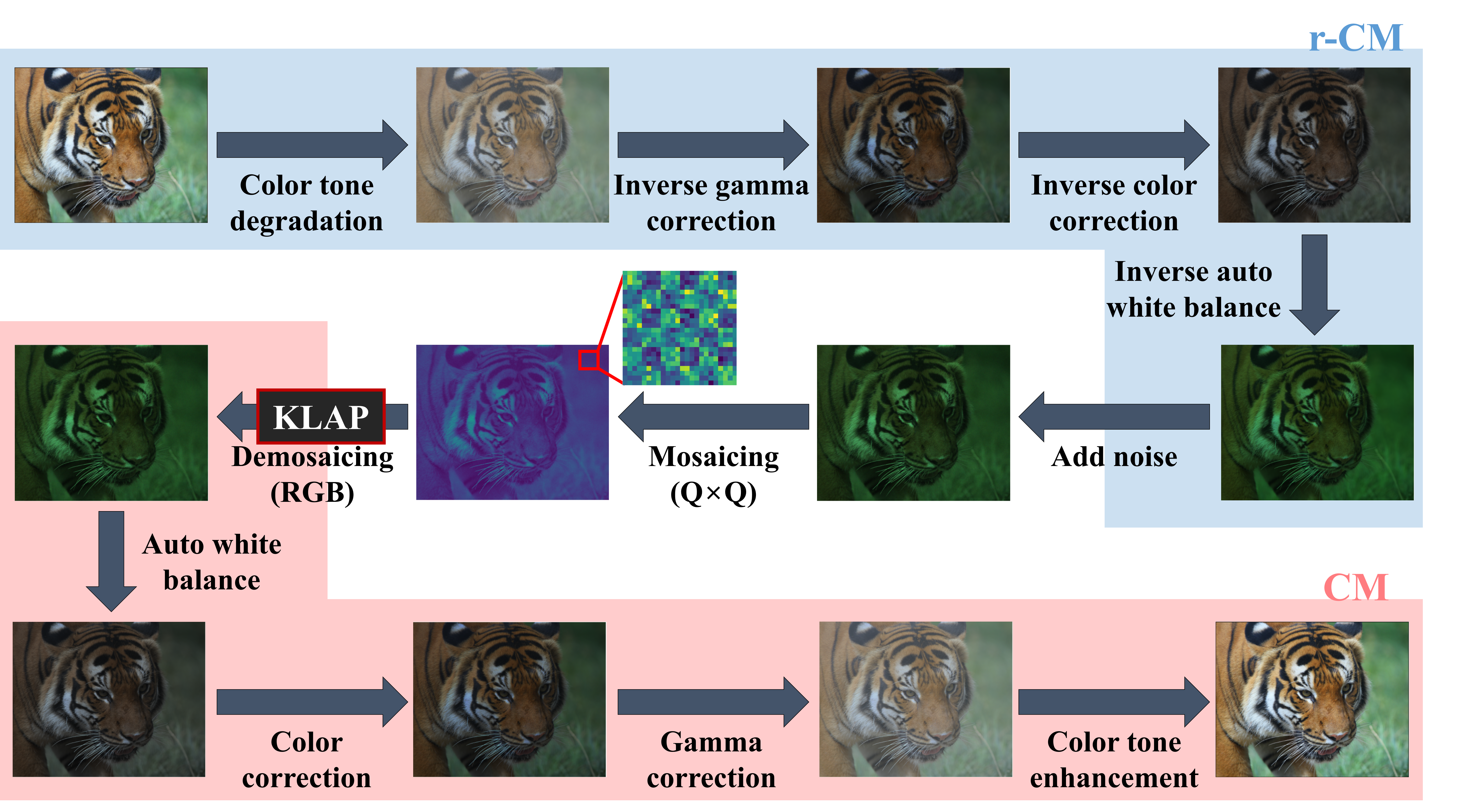}
    \captionof{figure}{Overview of our pipeline for synthesizing realistic RAW images, specifically for Q$\times$Q patterns.}
    \label{Supp_Fig1}
 
\end{center}%
}]
\let\thefootnote\relax\footnotetext{$*$  Equal contribution, $\dagger$ Corresponding author.}

\renewcommand{\thefigure}{S.\arabic{figure}}
\renewcommand{\thetable}{S.\arabic{table}}
\renewcommand{\thesection}{S.\arabic{section}}

\section{Detailed Data Synthesis for Demosaicing All CFAs}
\label{sec:Synthetic data}
As described in our paper, we generate synthetic ground truth (GT) by sequentially applying a 4-step \textbf{r}everse \textbf{C}olor-related \textbf{M}apping (r-CM) process. Then, we add mixed Poisson and Gaussian noise and performed mosaicing (\textit{i.e.}, CFA patterning) on the entire image to create synthetic RAW-like images (as shown in the blue shaded area in Fig.~\ref{Supp_Fig1}).
The r-CM process consists of the following modules: color tone degradation, inverse gamma correction, inverse color correction, and inverse auto white balance functions.
The color matrix (CM) is the inverse of the reverse color matrix (r-CM) and can only be applied to the output of the demosaicing (DM) model.

Note that we need to use r-CM for data synthesis on the open-source dataset to generate GT images, while CM can be ``optionally'' applied after demosaicing for better visualization in our paper.

\textbf{Color tone degradation.} Typically, color enhancement is performed in the latter part of the ISP chain. Therefore, we position the color tone degradation function at the beginning of r-CM. Inspired by ~\cite{brooks2019unprocessing}, we adopt a tone mapping function that uses a simple inverse smoothing curve, to perform color tone degradation on open-source dataset images in the r-CM process.
Note that the color tone enhancement function in CM is the inverse of color tone degradation in r-CM.

\textbf{Inverse gamma correction.} In the ISP chain, gamma correction is applied to image data to correct for the non-linear perception of brightness by the human eye. We use a gamma value setting of 2.2, which is standard for most cameras~\cite{ershov2022ntire, peng2022bokehme, xiong2012pixels, plotz2017benchmarking}.
In r-CM, the inverse function of gamma correction is applied, while in CM, standard gamma correction is performed.

\textbf{Inverse color correction.} We use a color correction function to adjust the colors captured by a camera's sensor to appear as they would to the human eye. The specific function we used is as follows:
\begin{equation*}
\begin{pmatrix}
R_{corrected} \\
G_{corrected} \\
B_{corrected}
\end{pmatrix}
= 
A
\begin{pmatrix}
R \\
G \\
B
\end{pmatrix},
\end{equation*}
where A is a 3x3 color correction matrix (CCM), which is applied to the pixel values (R, G, and B) to obtain the corrected RGB values ($R_{corrected}$, $G_{corrected}$, and $G_{corrected}$).
We obtain the CCM information from the CIS manufacturing company and apply it to our inverse color correction function after calculating the CCM's inverse.

\textbf{Inverse auto white balance.} We empirically adjusted the gains for R, G, and B channels in the auto white balance function to make white portions of the CIS RAW appear white as perceived by the human eye.
The inverse auto white balance in r-CM is obtained by reversing the values applied in the white balance module of the CM process.

\textbf{Noise synthesis.} We use the following practical mixed Poisson and Gaussian noise model~\cite{rim2022realistic,brooks2019unprocessing,xu2020noisy}:
\begin{equation}
\begin{split}
		x_n = \operatorname{Poisson}(\gamma y_n)/\gamma + \epsilon_n, \\
		\quad \epsilon \sim  \mathcal{N}(0,\sigma^2_\epsilon I),
		\quad n = 1, \ldots ,N, 
\end{split}
\end{equation}
where $y$ and $x$ are clean image and corrupted image, respectively. $\operatorname{Poisson}$ generates pixel intensity-dependent Poisson noise caused by photon sensing, $\gamma$ is a gain parameter which depends on the sensor and analog gain.
$\epsilon$ is signal independent Gaussian noise with standard deviation $\sigma$, and $N$ is the number of samples.
DF2K-CIS train and test datasets are generated using the following imaging parameters: $\gamma = 0.01$ and $\sigma = 0.02$. DF2K-CIS with strong noise test dataset are generated with parameters that are 4 times larger than those of DF2K-CIS: $\gamma = 0.04$ and $\sigma = 0.08$.

\section{Domain Gap Example: Inherent Grid Artifacts in CIS RAW}
\label{sec:GridArtifacts}

\begin{figure}[t]
    \centering
    \includegraphics[width=0.49\textwidth]{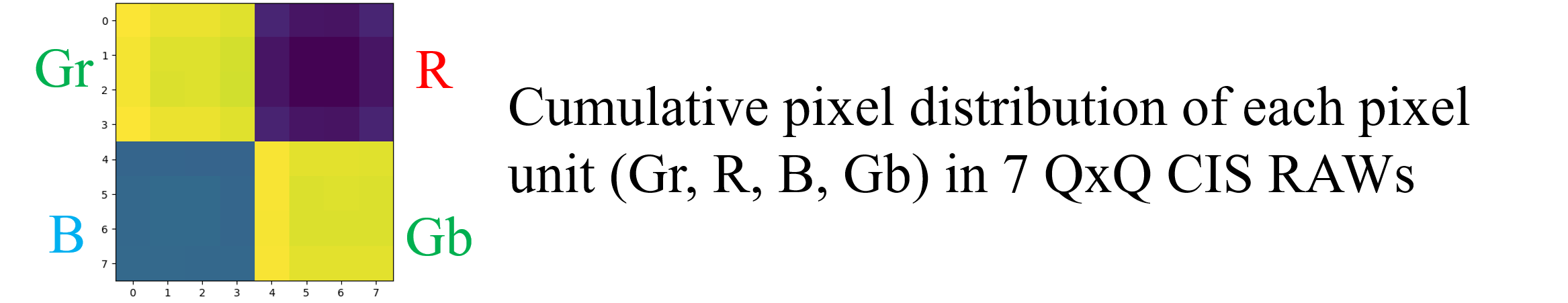}
    \caption{The cumulative pixel value distribution of each homogeneous pixel unit (Gr, R, B, and Gb) in 7 Q$\times$Q CIS RAW image samples. In our CIS RAW data, we observe a significant difference in signal values between inner and outer pixels in each Gr, R, B, and Gb pixel unit, which is mainly caused by crosstalk effect.}
    \label{Supp_Crosstalk}
\end{figure}

The differences in the distribution of pixels within each pixel unit are primarily caused by "crosstalk" effects, which result from mutual interference of each pixel signal in CIS hardware~\cite{khabir2021electrical, khabir2022characterization, li2022crosstalk}.
As shown in Fig.~\ref{Supp_Crosstalk}, in CIS QxQ RAW (before demosaicing), we observe that the signals in the center of each pixel unit, especially in the R channel, are stronger than those in the outer pixels, while the edges of each pixel unit, particularly the four corners, are weaker.
In addition to the cause of crosstalk phenomenon, the asymmetry between the inner and outer pixels in each pixel unit can vary across CIS devices, and this can manifest in various forms depending on the circuit configuration, component characteristics, product lines, and process capability of the CIS chip. The difference in pixel values in each of the homogeneous color units in CIS RAW may be causing grid artifacts.

\section{Adaptive Discriminative Filter-based Model for Specific CFA Pattern (ADP)}
\label{sec:ADP}

\subsection{Filter Attribution Integrated Gradients}
Xie \textit{et al.}~\cite{xie2021finding} propose FAIG, which identifies discriminative filters of specific degradation in blind super-resolution (SR) by computing integrated gradient (IG)~\cite{sundararajan2017axiomatic, sundararajan2016gradients} between the baseline and desired models.
In FAIG, the baseline model is denoted as $\theta_{from}$ and the model being updated is denoted as $\theta_{to}$ for each desired task. The function $\rho(\beta)$, where $\beta \in [0, 1]$, represents an uninterrupted straight line between the baseline and target models.
In that case, any certain route in $\rho(\beta)$ is represented by $\rho( \beta ) = \beta \theta_{from} + (1 - \beta) \theta_{to}$, where $\rho(1)=\theta_{from}$ and $\rho(0)=\theta_{to}$.
The FAIG on the continuous line space between two models is discretized as follows:
\begin{equation}
\begin{split}
\operatorname{FAIG}_i(\theta_{from}, \theta_{to}, x) \qquad\qquad\qquad\qquad\qquad\qquad \\
\approx  \left| \frac{1}{N}[\theta_{from}-\theta_{to}]_i \sum_{t=0}^{N-1}\left[\frac{\partial \mathcal{L}(\rho(\beta_t), x)}{\partial \rho(\beta_t)} \right]_i  \right|,
\end{split}
\label{eq:faig}
\end{equation}
where $N$ represents the total number of steps used in the integral approximation, and $N$ is set to 100 as in FAIG. $\beta_t$ and $i$ are $t/N$ and the kernel index, respectively.
We apply FAIG, originally proposed for denoising and deblurring, to multiple CFA sensor patterns in our demosaicing tasks.

\begin{table}[b]
\caption{Investigation of experiments according to KLAP (Ours) with filter location selection ratios (\textit{i.e.}, mask selection ratios in FAIG~\cite{xie2021finding}; $q$\%) in the DF2K-CIS test dataset. B.UM denotes the baseline UM. Note that Avg. and Par. denotes mean of all CFAs' PSNR (dB) and the number of parameters (M).
}
\label{tab:filterLocationSelection}
\centering
\begin{adjustbox}{width=0.90\linewidth}
\begin{tabular}{cc|cccc|cc}
\hline
Method & $q$ & Ba. & Qu. & No. & QxQ & Avg. & Par. \\ \hline\hline
B.UM & 0  & 41.90  &  41.40 &  41.03 &  41.09 & 41.35  & 17.1 \\ \hline
KLAP&0.1 &  42.16 &  41.50  &  41.16 &  41.19  & 41.50  & 17.2  \\
KLAP&0.5 &  42.20 &  41.71  &  41.38 &  41.38 & 41.67  & 17.4  \\
KLAP&1 &  42.25 &  41.75  &  41.42 &  41.41 & 41.71  & 17.8  \\
KLAP&3 & 42.31  &  41.80 & 41.46  &  41.45 & 41.75  & 19.2  \\
KLAP&5 &  42.34 &  41.82  & 41.49  & 41.48  & 41.78  &  20.5  \\ 
KLAP&10 &  42.38 &  41.88  & 41.55  & 41.53  & 41.83  &  23.9  \\ 
KLAP&15 &  42.41 &  41.92  & 41.59  & 41.58  & 41.87  &  27.4  \\ \hline
\end{tabular}
\end{adjustbox}
\end{table}

\begin{figure}[t]
    \centering
    \includegraphics[width=0.48\textwidth]{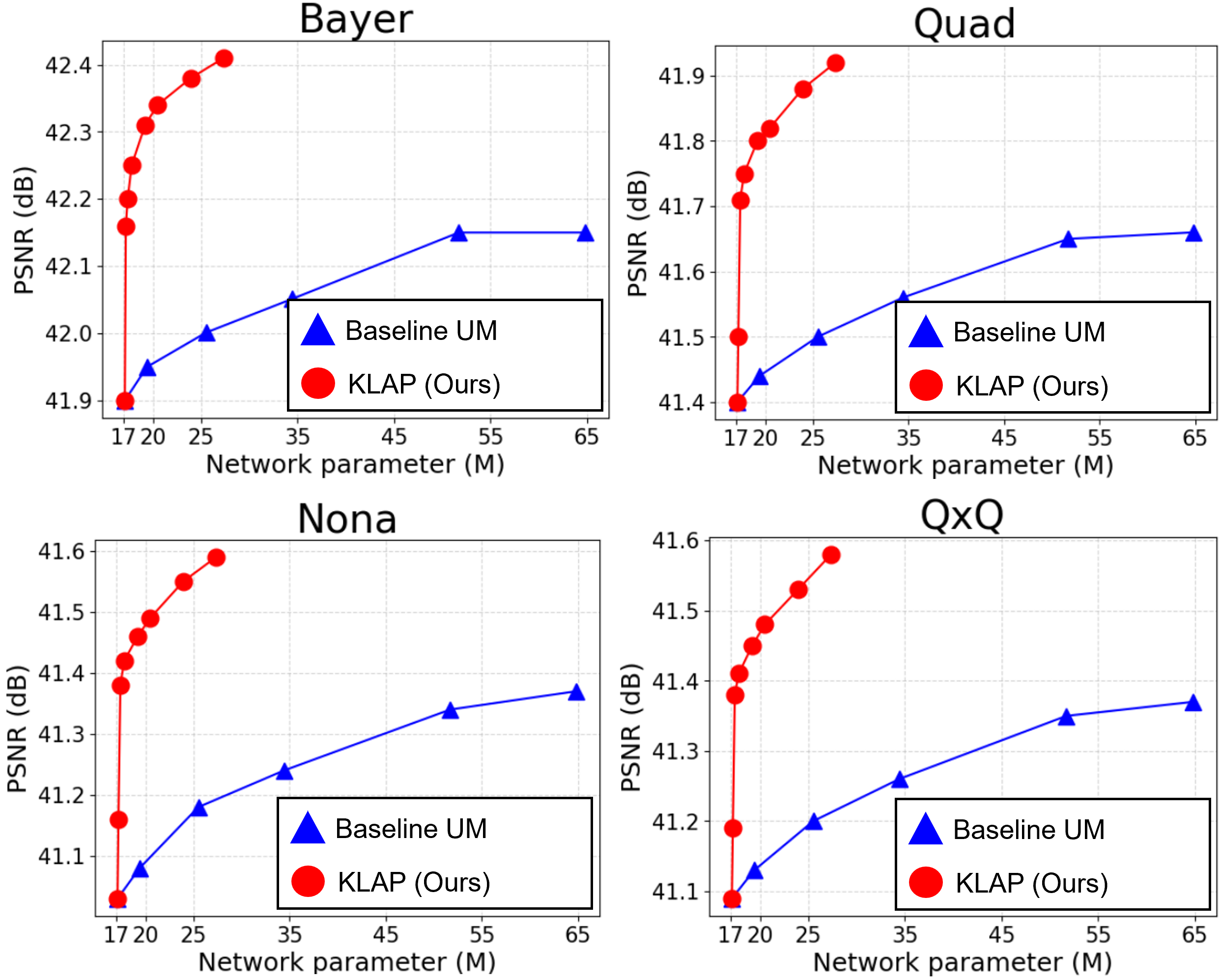}
    \caption{Performance comparisons between Baseline UM with increased network sizes (17.1M, 19.4M, 25.5M, 34.4M, 51.7M, and 64.8M) and KLAP (Ours) with mask ratios $q\%$ ($0\%$, $0.1\%$, $0.5\%$, $1\%$, $3\%$, $5\%$, $10\%$ and $15\%$, respectively) on DF2K-CIS test dataset. The network size of KLAP (Ours) with mask ratios of $q\%$ (0\%, 0.1\%, 0.5\%, 1\%, 3\%, 5\%, 10\%, and 15\%) are 17.1M, 17.2M, 17.8M, 17.8M, 19.2M, 20.5M, 23.9M, and 27.4M, respectively. Our approach produces significantly higher performance results even with 3.5 times larger Baseline UM method.}
    \label{Supp_Fig3}
\end{figure}

\begin{figure*}[t]
    \centering
    \includegraphics[width=0.85\textwidth]{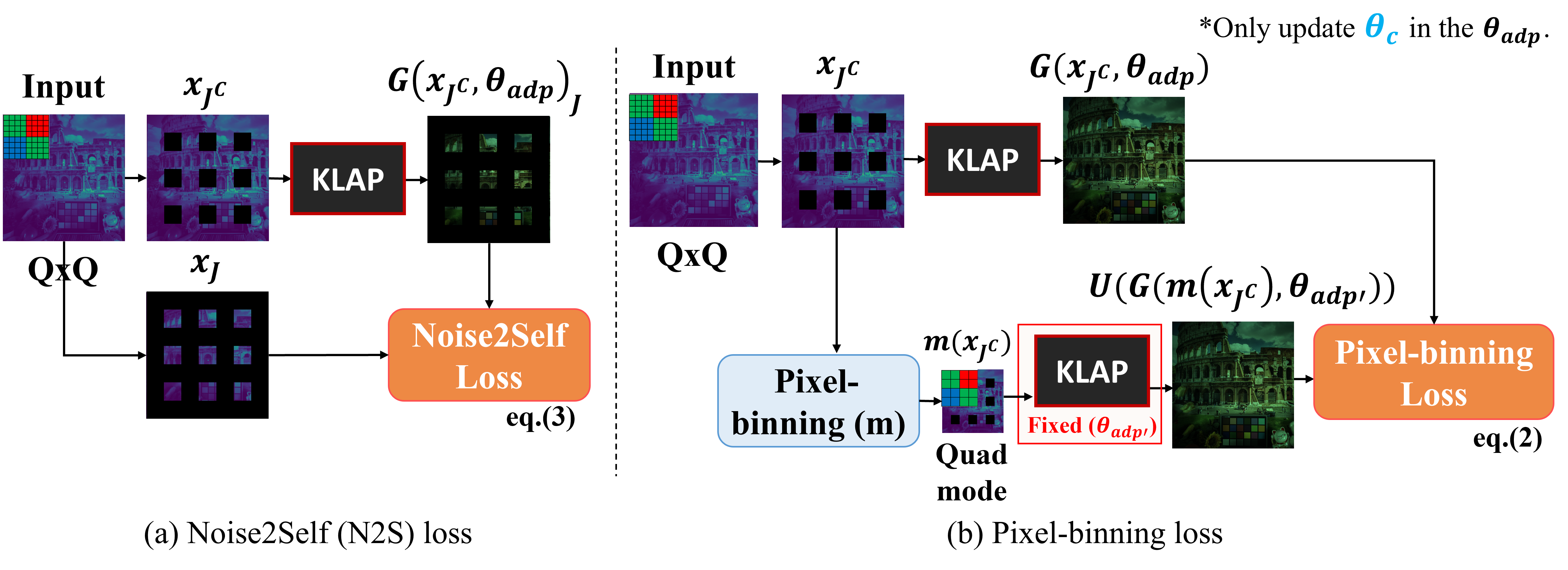}
    \caption{The specific processes for calculating 2 loss functions in our proposed method, KLAP-M, which is KLAP with meta-test learning: (a) Noise2Self (N2S) loss, and (b) Pixel-binning loss. 
    }
    \label{Supp_Fig2}
\end{figure*}

\subsection{Mask Ratio of FAIG in ADP}
We choose a mask ratio ($q$) as 1\% in ADP for each CFA in our KLAP framework, to balance demosaicing performance and efficiency (as shown in Sec. 4.2 and Fig. 4(b)).
Increasing $q$ improves performance but with diminishing returns and increased parameters (Tab.~\ref{tab:filterLocationSelection}). Compared to Baseline-UM (B.UM), our proposed method using mask ratio 1\% for all 4 demosaicing types requires an additional 4\% of network parameters.

Furthermore, our KLAP achieves significantly better results even when increasing the size of the Baseline UM method by 3.5 times, as shown in Fig.~\ref{Supp_Fig3}.

\section{Meta-learning during Inference}
\label{sec:Meta-test learning}

\subsection{Definition of the term ``meta-test''}
In our paper, we name the process of fine-tuning with meta-learning during inference as meta-test learning in KLAP-M. The term ``meta-test'' typically refers to the process of improving performance on various generalization scenarios with only a few trials on unseen data~\cite{yang2021joint,finn2018probabilistic,zhao2021learning,ni2022meta,pong2022offline,rivera2022visual}.
In general, the meta-test process works in conjunction with the meta-training process. The meta-training process optimizes the model to improve the accuracy of meta-test samples using source data.
We use ADP in the second step of our KLAP framework to only adjust the important kernel ($\theta_c$) for each CFA demosaicing during training in order to improve the accuracy of meta-test. This can be seen as a type of meta-training process. 
In our paper, we define the process of fine-tuning only $\theta_c$ in KLAP during model inference as meta-test learning to achieve robust results even for undefined artifacts caused by CIS device features and shooting environments.

\subsection{Noise2Self and Pixel Binning Loss.}
To aid in a more thorough understanding in our meta-test learning process, KLAP-M, we provide a more detailed explanation of Noise2Self (N2S)~\cite{batson2019noise2self} loss and pixel-binning loss in Fig.~\ref{Supp_Fig2} (a) and (b).

\textbf{Noise2Self (N2S) loss.}
We choose Noise2Self~\cite{batson2019noise2self} among many self-supervised denoising methods~\cite{zhussip2019training,zhussip2019extending,lehtinen2018noise2noise,huang2021neighbor2neighbor,batson2019noise2self,byun2021fbi}.
To calculate N2S loss, the L1 loss is computed between an output of an image inputted into KLAP, where the empty pixels of $x_{J^c}$ are interpolated, and $x_J$.
In our study, we utilize the same masking scheme for each $J$ as outlined in the N2S paper~\cite{batson2019noise2self}. Each $J$ samples a single pixel selected within each 4$\times$4 window (\textit{i.e.}, 6.25\% of the number of pixels in each image).
In the original N2S method, the interpolation function for $x_{J^c}$ use a 3$\times$3 kernel to compute the average value of the surrounding pixels for interpolation. However, we consider the characteristics of the RGB channel and calculate the average value of the surrounding values corresponding to that channel for interpolation.
In the case of Bayer, we set a size of window to 6$\times$6 and use 5$\times$5 kernel for interpolation to prevent overlap.
The use of N2S loss term has the effect of removing independent noise.

\textbf{Pixel-binning loss.}
As mentioned in Sec. 3.1 in our paper, pixel binning is applied differently depending on the input pattern status of the CFA. Similarly, the proposed pixel binning loss based on CIS domain knowledge is also applied differently according to the CFA pattern.
When using the average-based pixel binning operation ($m$), the Q$\times$Q CFA pattern is converted to Quad or Bayer pattern. Nona and Quad patterns are converted to Bayer pattern.
Note that pixel binning operation ($m$) does not exist in the Bayer pattern.
The upsampling operation ($U$) employs a bilinear function to restore the original resolution, which may have been altered due to the pixel binning operation ($m$).

\section{Implementation Details}
\label{sec:Meta-test learning}

In our experiments, we use a patch size of 240$\times$240 to cover all of Bayer, Quad, Nona, and Q$\times$Q CFAs. The model is trained using the ADAM optimizer with a batch size of 32 and an initial learning rate of $2\times10^{-4}$. We apply the cosine annealing learning rate decay technique with a minimum learning rate of $1\times 10^{-6}$.
To ensure a fair comparison, we evaluate the proposed method on the same conditions with an NVIDIA A6000 GPU using PyTorch~\cite{paszke2019pytorch}. Our architecture is similar to the NAFNet~\cite{chen2022simple} architecture, which has state-of-the-art performance in IM-based image restoration. We use the official codes provided by the authors of Chen~\cite{chen2022learning} and Li~\cite{li2022all}.
In the Table 1, TKL denotes the method of applying NAFNet-based TKL, and Chen denotes the method of applying MBSDN-based TKL.

\begin{figure*}[]
    \centering
    \includegraphics[width=1.0\textwidth]{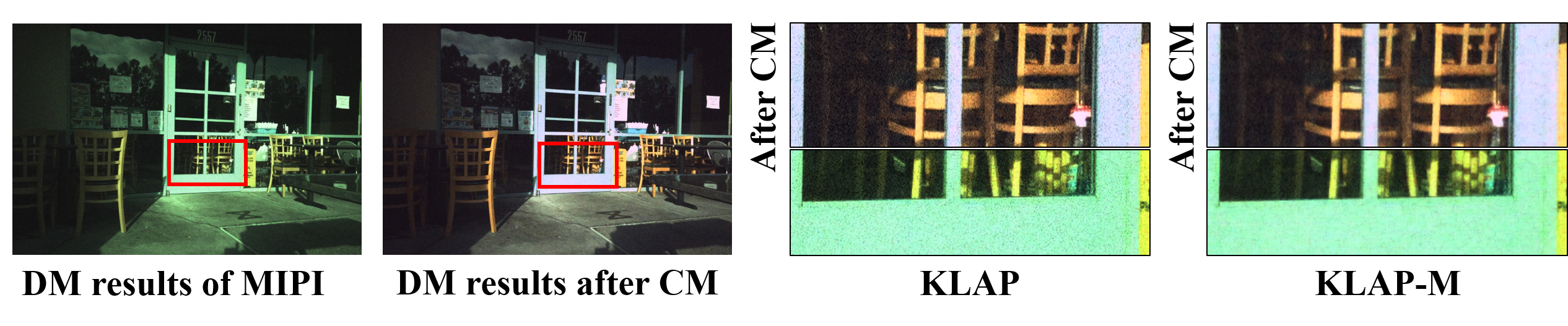}
    \caption{{Results of KLAP and KLAP-M on MIPI `22 Quad.}}
    \label{rebut_MIPI_inference}
    \label{Supp_Fig_mipi}
\end{figure*}

\begin{figure*}[!ht]
\centering
\includegraphics[width=1.0\textwidth]{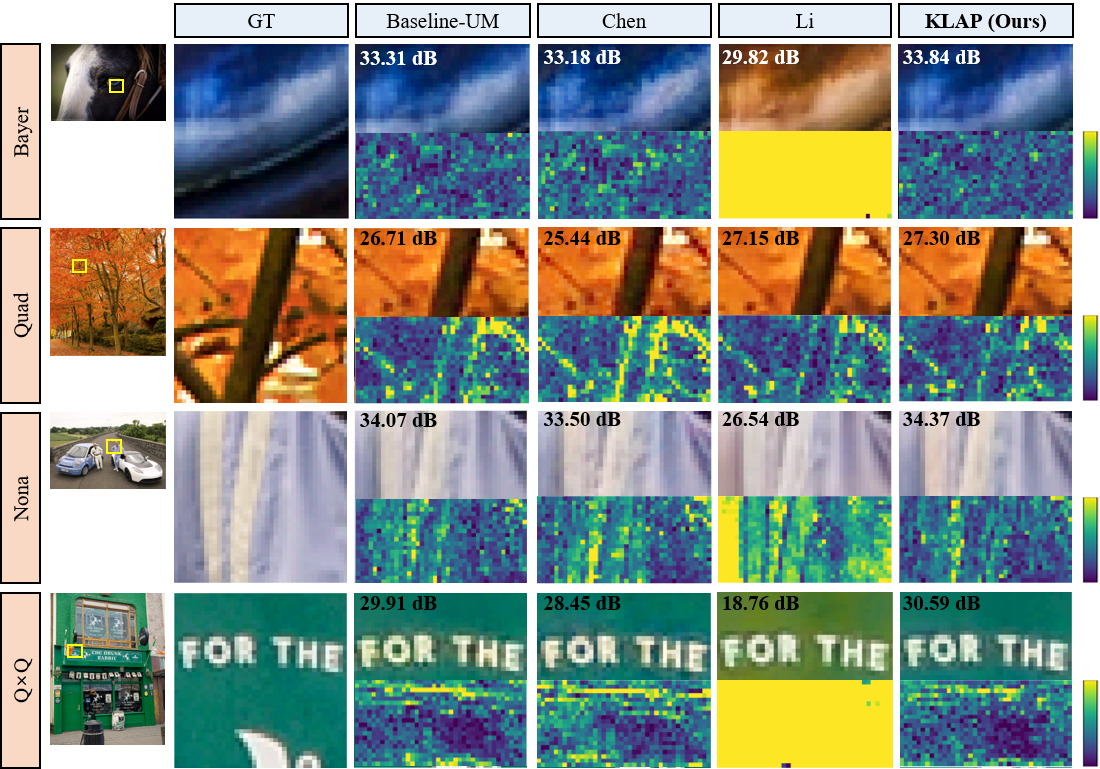}
    \caption{
    Comparisons of demosaiced images {\bfseries(top)} from different methods and their difference maps {\bfseries(bottom)} on the synthetic RAW (DF2K-CIS) test dataset. The PSNR (dB) values displayed in the top-left corner of each image are calculated using the entire image.
    }
    \label{Supp_Fig4}
\end{figure*}

\begin{figure*}[!ht]
\centering
\includegraphics[width=1.0\textwidth]{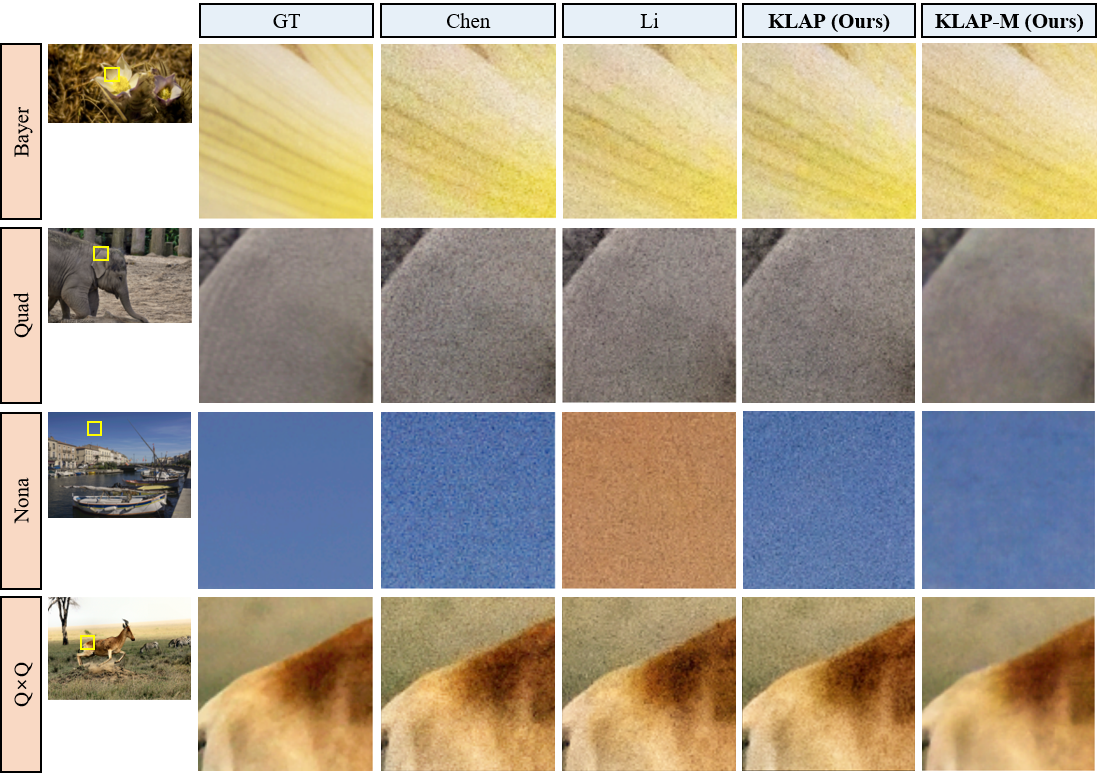}
    \caption{
    Comparisons among different methods of robustness on DF2K-CIS with strong noise test dataset. The noise parameters used in the test are four times larger than the noise parameters used in the training. The number of meta-learning iterations in KLAP-M is set to 10, based on empirical determination through visualization of outputs in our experiments.
    }
    \label{Supp_Fig5}
\end{figure*}

\section{Demosaicing Methods Comparison}
As the pioneers in integrated DM tasks for various sensor CFAs, we compared our proposed KLAP method with state-of-the-art integrated image restoration methods~\cite{chen2022learning,li2022all} due to the lack of existing unified DM research.
We conduct experiments on recent DM methods~\cite{cho2022pynet,kim2021recent} for specific sensors, as well as ~\cite{wang2022uformer}, one of the winners in the MIPI `22~\cite{yang2022mipi} competition. The results, as shown in Tab.~\ref{Supp_Tab_S.2} indicates that none surpassed ours.

\begin{table}[] 
\centering 
\caption{Performance comparisons with existing DM methods for specific sensors in IMs}
\begin{adjustbox}{width=0.9\linewidth} 
\begin{tabular}{c|cccc|c}
\hline
Method & Bayer & Quad & Nona & QxQ & Par. \\ \hline\hline
IM~\cite{cho2022pynet} & 37.03 & 37.38 & 36.65 & 36.44 & 4.2\\
IM~\cite{kim2021recent} & 41.33 & 40.81 & 39.85 & 37.02& 13.8\\
IM~\cite{yang2022mipi} & 41.89 & 41.19 & 40.60 & 40.74& 83.0\\ \hline
KLAP & \textbf{42.25}  & \textbf{41.75} & \textbf{41.42} & \textbf{41.41} & \textbf{17.8} \\ \hline
\end{tabular}
\end{adjustbox}
\vspace{-0.3cm}
\label{Supp_Tab_S.2}
\end{table}

\section{Additional RAW Evaluation}
MIPI `22 competition~\cite{yang2022mipi} emphasizes Quad-to-Bayer \textit{re-mosaicing}, not demosaicing, so the definition of ground truth (GT) differs from our research focus.
Nevertheless, we conducted inference on the MIPI inputs using KLAP and KLAP-M, as shown in Fig.~\ref{Supp_Fig_mipi}, effectively reducing visual artifacts and validating their performance.
The MIPI challenge uses synthetic inputs without RGB GT, emphasizing the challenges of acquiring real CIS RAW data. This highlights the importance of our self-supervised learning approach for \textit{real} sensor RAW in real-world scenarios.

\section{Results}
\begin{figure*}[!ht]
\centering
\includegraphics[width=0.8\textwidth]{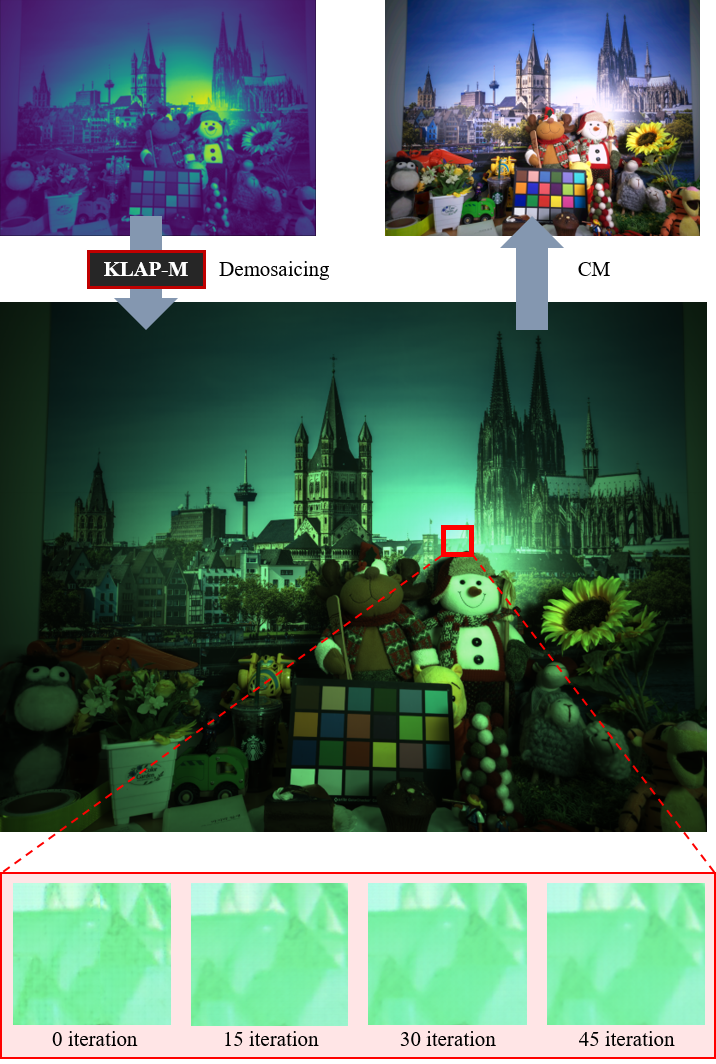}
    \caption{
    The demosaiced output images of Q$\times$Q CIS RAW (48MP) with various iterations of KLAP-M.
    }
    \label{Supp_Fig6}
\end{figure*}

Fig.~\ref{Supp_Fig4}  illustrates the qualitative results of  the Baseline-UM (2nd column) with NAFNet, existing methods ( Chen (3rd column),  Li (4th column)) and our proposed KLAP (5th column), evaluated on the synthetic RAW (DF2K-CIS) test dataset. 
Fig.~\ref{Supp_Fig5} presents the qualitative results of  prior arts (Chen (2nd column), Li (3rd column)) and our proposed KLAP (4th column) and KLAP-M (5th column) on the synthetic RAW (DF2K-CIS) with strong noise test dataset. 
Our proposed KLAP method visually outperform other state-of-the-art methods on DK2K-CIS test dataset, and our proposed KLAP-M method shows visually superior results compared to other state-of-the-art methods on the DK2K-CIS test with strong noise dataset, thanks to meta-learning during inference.
Fig.~\ref{Supp_Fig6} shows our proposed KLAP-M inference output on the real CIS RAW data.
Without meta-learning applied (0 iteration) in Fig.~\ref{Supp_Fig6}, artifacts exist. On the other hand, as the number of meta-learning iterations increases, the artifacts gradually disappear. We selected 45 iterations for real data using this method. Furthermore, we observed that similar results were obtained even with further increases in iterations.

\begin{figure*}[!ht]
\centering
\includegraphics[width=0.9\textwidth]{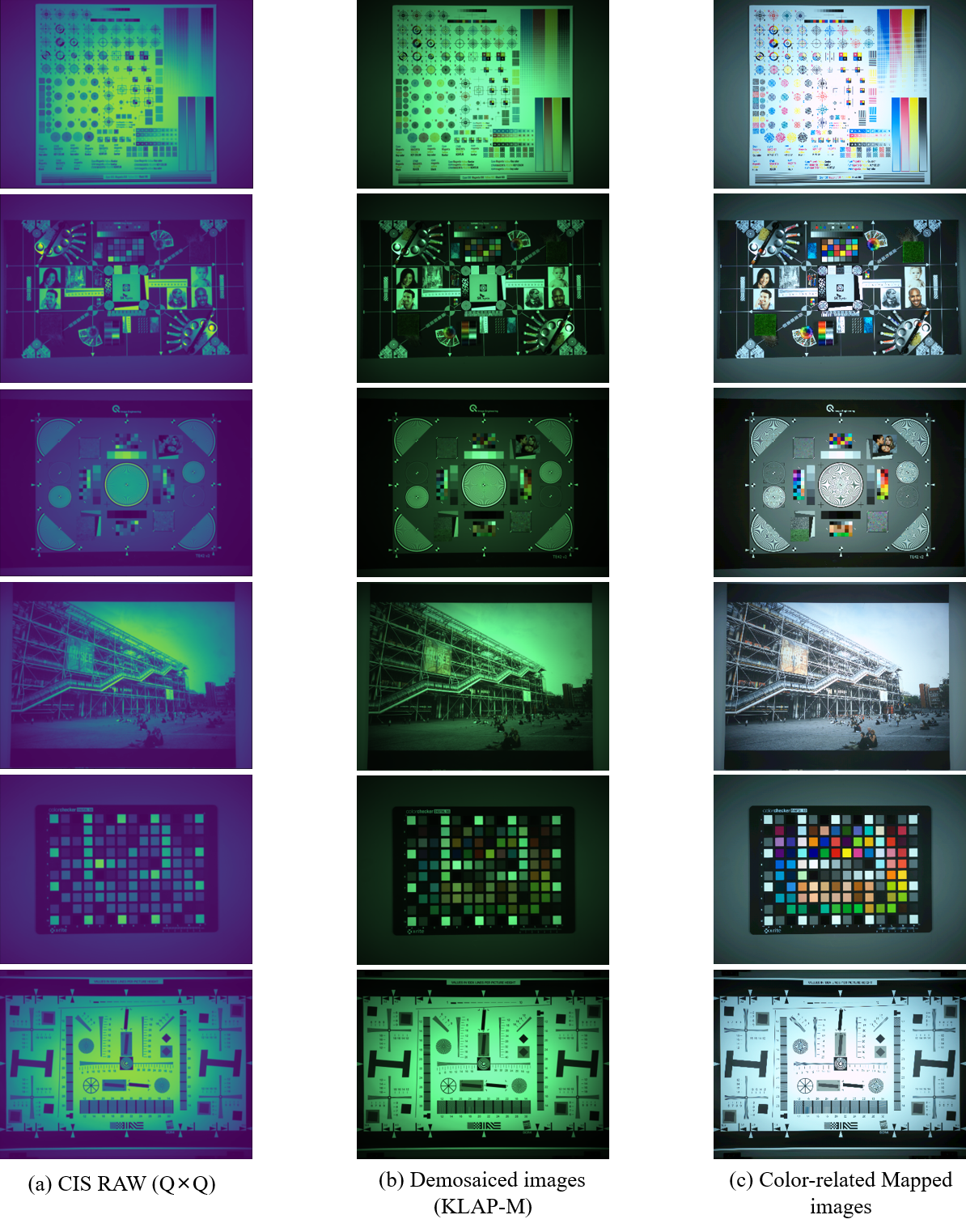}
    \caption{
    Additional images of CIS RAW data. (a) CIS Q$\times$Q RAW data, (b) demosaiced output images obtained using KLAP-M inference, and (c) the same images as in (b) after applying CM (Color-related Mapping function). Note that in (c), it can be perceptually observed that CM works well not only on synthetic RAW images but also on real CIS RAW images.
    }
    \label{Supp_Fig7}
\end{figure*}

Fig.~\ref{Supp_Fig7} shows our proposed KLAP-M inference output on the real CIS RAW data set.

\end{document}